\documentclass[12pt,english,floatfix,nofootinbib,superscriptaddress,aps,prd,preprint]{revtex4}

\usepackage[utf8]{inputenc}
\usepackage{float}
\usepackage{array}
\usepackage{bbold}
\usepackage{lipsum}
\usepackage{dsfont}
\usepackage{graphicx}
\usepackage{amsmath,amsthm,amsfonts,amssymb}
\usepackage{graphicx}
\usepackage[english]{babel} 
\usepackage{color}
\usepackage{tensor}
\usepackage{esint}
\usepackage[dvips]{epsfig}
\usepackage[dvips]{graphicx}
\usepackage{float}
\usepackage{units}
\usepackage{textcomp}
\usepackage{mathrsfs}
\usepackage{amsmath}
\usepackage[makeroom]{cancel}
\usepackage{amssymb}
\usepackage{amsbsy}
\usepackage{amsfonts}
\usepackage{amssymb,mathrsfs,xcolor}
\usepackage{esint}
\usepackage{braket}
\usepackage{array}
\usepackage{graphicx}

\usepackage{wasysym}
\usepackage{multirow}
\usepackage{wrapfig}
\usepackage{subfig}

\usepackage{stmaryrd}
\usepackage{upgreek}

\makeatletter

\makeatletter\usepackage{babel}

\usepackage{hyperref}
\hypersetup{
    colorlinks,
    citecolor=blue,
    filecolor=green,
    linkcolor=purple,
    urlcolor=red,
}

\usepackage{slashed}

\newcommand{\ie}{\begin{equation}}
\newcommand{\fe}{\end{equation}}
\newcommand{\se}{\begin{eqnarray}}
\newcommand{\ff}{\end{eqnarray}}

\begin{document}

\title{Remarks on a nonlinear electromagnetic extension in AdS Reissner--Nordström spacetime}

\author{A. A. Ara\'{u}jo Filho}
\email{dilto@fisica.ufc.br}

\affiliation{Departamento de Física, Universidade Federal da Paraíba, Caixa Postal 5008, 58051-970, João Pessoa, Paraíba,  Brazil.}


\date{\today}

\begin{abstract}

We explore the gravitational properties of a nonlinear electromagnetic extension of an AdS Reissner--Nordström black hole. Our study begins with an analysis of the metric function and horizon structure, followed by calculations of the Ricci and Kretschmann scalars and an evaluation of the non--vanishing Christoffel symbols. These calculations allow us to examine geodesics and their influence on the photon sphere, and shadow formation. In the thermodynamic framework, we evaluate essential quantities, including the Hawking temperature, entropy, heat capacity, Gibbs free energy, and Hawking radiation emission. We further investigate black hole evaporation by estimating the evaporation timescale as the black hole approaches its final state. Additionally, quasinormal modes for scalar and vector perturbations are computed using the WKB approximation to characterize oscillatory behavior of the system. Finally, a time--domain analysis is provided in order to examine the evolution of these perturbations.

\end{abstract}

\maketitle


\section{Introduction}

Highly magnetized compact objects, such as magnetars and neutron stars, demonstrate nonlinear electromagnetic (NLE) effects that call for revisions to traditional Maxwell theory \cite{bronnikov2001regular,mosquera2004non,denisov2017vacuum,javed2019deflection,soares2023thermodynamics,sorokin2022introductory,denisova2019compact,lyutikov2011electromagnetic,denisov2016pulsar,torres2011unified,lobo2007gravastars}. The Einstein--NLE equations provide solutions that clarify the physics surrounding strongly magnetized black holes and offer practical models for testing computational simulations. Moreover, stationary configurations with NLE fields provide a deeper view of rotating astrophysical bodies, particularly in relation to the issue of singularities. Numerous static black hole solutions with NLE sources avoid these singularities \cite{NEL03}, and a notable advancement includes a Kerr--Newman black hole modified by Euler--Heisenberg nonlinearities \cite{NEL04}.

Numerous NLE Lagrangians, formulated as nonlinear expressions of electromagnetic invariants, offer promising frameworks for extending the traditional Ker--Newman solution \cite{burinskii2014kerr,hassaine2008higher, galindo2024nonlinear,dymnikova2015electromagnetic,ayon2024unveiling,kubizvnak2022slowly,lammerzahl2018slowly, burinskii1999super,dymnikova2015regular, burinskii2008dirac,garcia2021stationary, hintz2018non,zilhao2014testing}. Recent research has yielded exact solutions to the Einstein--NLE field equations that describe rotating black holes with parameters including mass, angular momentum, electric charge, a cosmological constant, and a nonlinear electromagnetic parameter \cite{NEL06,NEL05}. Within this framework, theories that maintain Lorentz and gauge invariance have been comprehensively examined, beginning with Plebański’s foundational work \cite{NEL07} and further developed by Boillat \cite{NEL08}. These theories profoundly impact the understanding of light propagation, as they predict that light rays trace null geodesics associated with two distinct optical metrics. Subsequent investigations by Novello and others \cite{NEL10} reexamined these findings, and later work by Obukhov and Rubilar \cite{NEL11} showed that in certain NLE theories, the Fresnel equation governing wave covectors decomposes, leading to a remarkable phenomenon: the birefringence.

In Ref. \cite{galindo2024nonlinear}, the authors introduced a nonlinear electromagnetic extension of the Kerr–Newman solution with a cosmological constant, along with a brief analysis of its static limit. More recently, a study investigated a specific case of this solution with \(\Lambda \to 0\) \cite{AraujoFilho:2024xhm}, examining gravitational characteristics (see also a recently exact solution to the Klein--Gordon equation in the most general Kerr type black hole, i.e. the Dyonic Kerr Sen  \cite{senjaya2024exact}, the Kerr--Newmann \cite{Vieira:2021nha} and also the Schwarzschild AdS \cite{senjaya2024scalar}) by analyzing light trajectories, thermodynamic aspects, and quasinormal modes.

To date, a thorough examination of the case with a nonzero cosmological constant (\(\Lambda \neq 0\)) remains absent in the literature. Addressing this gap, we investigate the gravitational characteristics of such a system. We begin with an analysis of the metric function \(f(r)\) and the associated horizons, followed by computations of both the Ricci and Kretschmann scalars and an evaluation of the non--vanishing Christoffel symbols. With these, we determine the geodesics, examining their impact on the photon sphere, and shadow formation. In the thermodynamic context, we assess several key quantities: the Hawking temperature, entropy, heat capacity, Gibbs free energy, and Hawking radiation emission. We further explore black hole evaporation, estimating the evaporation timescale as the black hole reaches its final state. Additionally, we calculate quasinormal modes for scalar and vector perturbations using the WKB approximation to capture oscillatory dynamics of the system. Finally, we conduct an analysis of the time--domain solution to understand its implications on the evolution of perturbations.


\section{The general features of the black hole solution}

In the absence of rotational effects, the solution simplifies to a static form, which describes a nonlinear electrodynamics modification of the Reissner--Nordström metric \cite{galindo2024nonlinear}
\ie
\begin{split}
\mathrm{d}s^{2} = & - \left[  1 - \frac{2M}{r} + \frac{\mathfrak{Q}}{r^{2}}(1 + \xi r^{3})  + \frac{\Lambda}{3}r^{2}  \right] \mathrm{d} t^{2} + \frac{1}{\left[  1 - \frac{2M}{r} + \frac{\mathfrak{Q}}{r^{2}}(1 + \xi r^{3}) + \frac{\Lambda}{3}r^{2}    \right]} \mathrm{d}r^{2} \\
& + r^{2} \mathrm{d}\theta^{2} + r^{2} \sin^{2}\theta \mathrm{d}\phi^{2}.
\end{split}
\fe

Here, \( \mathfrak{Q} \equiv Q^{2}_{e} + Q^{2}_{m} \) represents the effective charge, where \( Q_{e} \) and \( Q_{m} \) refer to the electric and magnetic charges, respectively, and \(\Lambda\) is the cosmological constant. The metric function \( f(r) \) is given by \( f(r) = g(r) \equiv 1 - \frac{2M}{r} + \frac{\mathfrak{Q}}{r^{2}}(1 + \xi r^{3}) + \frac{\Lambda}{3}r^{2} \). Notice that \( f(r) \) can be expressed as $
f(r) = \frac{\frac{\Lambda}{3}r^4 + \mathfrak{Q}\xi r^3 + r^2 - 2Mr + \mathfrak{Q}}{r^2}, $
making it evident that the numerator is a quartic polynomial. In this manner, solving \( f(r) = 0 \) yields four distinct horizon solutions, each naturally expressed as a function of the parameters \(\mathfrak{Q}\), \(M\), and \(\xi\)
\ie
\begin{split}
\label{eventttt}
&  r_{h} =  +\frac{1}{2} \sqrt{\frac{\sqrt[3]{\beta }}{3 \sqrt[3]{2} \Lambda }-\frac{2}{\Lambda }+\frac{3 \sqrt[3]{2} (6 M \xi  \mathfrak{Q}+4 \Lambda  \mathfrak{Q}+1)}{\sqrt[3]{\beta } \Lambda }+\frac{9 \xi ^2 \mathfrak{Q}^2}{4 \Lambda ^2}}-\frac{3 \xi  \mathfrak{Q}}{4 \Lambda } \\ & - \frac{1}{2} \left[-\frac{\sqrt[3]{\beta }}{3 \sqrt[3]{2} \Lambda }-\frac{4}{\Lambda }-\frac{\frac{48 M}{\Lambda }-\frac{27 \xi ^3 \mathfrak{Q}^3}{\Lambda ^3}+\frac{36 \xi  \mathfrak{Q}}{\Lambda ^2}}{4 \sqrt{\frac{\sqrt[3]{\beta }}{3 \sqrt[3]{2} \Lambda }-\frac{2}{\Lambda }+\frac{3 \sqrt[3]{2} (6 M \xi  \mathfrak{Q}+4 \Lambda  \mathfrak{Q}+1)}{\sqrt[3]{\beta } \Lambda }+\frac{9 \xi ^2 \mathfrak{Q}^2}{4 \Lambda ^2}}} \right. \\
& \left. -\frac{3 \sqrt[3]{2} (6 M \xi  \mathfrak{Q}+4 \Lambda  \mathfrak{Q}+1)}{\sqrt[3]{\beta } \Lambda }+\frac{9 \xi ^2 \mathfrak{Q}^2}{2 \Lambda ^2}\right]^{1/2},
\end{split}
\fe
\ie
\begin{split}
& r_{1} = -\frac{1}{2} \sqrt{\frac{\sqrt[3]{\beta }}{3 \sqrt[3]{2} \Lambda }-\frac{2}{\Lambda }+\frac{3 \sqrt[3]{2} (6 M \xi  \mathfrak{Q}+4 \Lambda  \mathfrak{Q}+1)}{\sqrt[3]{\beta } \Lambda }+\frac{9 \xi ^2 \mathfrak{Q}^2}{4 \Lambda ^2}}-\frac{3 \xi  \mathfrak{Q}}{4 \Lambda } \\
& + \frac{1}{2} \left[-\frac{\sqrt[3]{\beta }}{3 \sqrt[3]{2} \Lambda }-\frac{4}{\Lambda }-\frac{\frac{48 M}{\Lambda }-\frac{27 \xi ^3 \mathfrak{Q}^3}{\Lambda ^3}+\frac{36 \xi  \mathfrak{Q}}{\Lambda ^2}}{4 \sqrt{\frac{\sqrt[3]{\beta }}{3 \sqrt[3]{2} \Lambda }-\frac{2}{\Lambda }+\frac{3 \sqrt[3]{2} (6 M \xi  \mathfrak{Q}+4 \Lambda  \mathfrak{Q}+1)}{\sqrt[3]{\beta } \Lambda }+\frac{9 \xi ^2 \mathfrak{Q}^2}{4 \Lambda ^2}}} \right. \\
& \left. -\frac{3 \sqrt[3]{2} (6 M \xi  \mathfrak{Q}+4 \Lambda  \mathfrak{Q}+1)}{\sqrt[3]{\beta } \Lambda }+\frac{9 \xi ^2 \mathfrak{Q}^2}{2 \Lambda ^2}\right]^{1/2},
\end{split}
\fe
\ie
\begin{split}
&  r_{2} =  -\frac{1}{2} \sqrt{\frac{\sqrt[3]{\beta }}{3 \sqrt[3]{2} \Lambda }-\frac{2}{\Lambda }+\frac{3 \sqrt[3]{2} (6 M \xi  \mathfrak{Q}+4 \Lambda  \mathfrak{Q}+1)}{\sqrt[3]{\beta } \Lambda }+\frac{9 \xi ^2 \mathfrak{Q}^2}{4 \Lambda ^2}}-\frac{3 \xi  \mathfrak{Q}}{4 \Lambda } \\ & - \frac{1}{2} \left[-\frac{\sqrt[3]{\beta }}{3 \sqrt[3]{2} \Lambda }-\frac{4}{\Lambda }-\frac{\frac{48 M}{\Lambda }-\frac{27 \xi ^3 \mathfrak{Q}^3}{\Lambda ^3}+\frac{36 \xi  \mathfrak{Q}}{\Lambda ^2}}{4 \sqrt{\frac{\sqrt[3]{\beta }}{3 \sqrt[3]{2} \Lambda }-\frac{2}{\Lambda }+\frac{3 \sqrt[3]{2} (6 M \xi  \mathfrak{Q}+4 \Lambda  \mathfrak{Q}+1)}{\sqrt[3]{\beta } \Lambda }+\frac{9 \xi ^2 \mathfrak{Q}^2}{4 \Lambda ^2}}} \right. \\
& \left. -\frac{3 \sqrt[3]{2} (6 M \xi  \mathfrak{Q}+4 \Lambda  \mathfrak{Q}+1)}{\sqrt[3]{\beta } \Lambda }+\frac{9 \xi ^2 \mathfrak{Q}^2}{2 \Lambda ^2}\right]^{1/2},
\end{split}
\fe
and
\ie
\begin{split}
&  r_{3} =  +\frac{1}{2} \sqrt{\frac{\sqrt[3]{\beta }}{3 \sqrt[3]{2} \Lambda }-\frac{2}{\Lambda }+\frac{3 \sqrt[3]{2} (6 M \xi  \mathfrak{Q}+4 \Lambda  \mathfrak{Q}+1)}{\sqrt[3]{\beta } \Lambda }+\frac{9 \xi ^2 \mathfrak{Q}^2}{4 \Lambda ^2}}-\frac{3 \xi  \mathfrak{Q}}{4 \Lambda } \\ & + \frac{1}{2} \left[-\frac{\sqrt[3]{\beta }}{3 \sqrt[3]{2} \Lambda }-\frac{4}{\Lambda }-\frac{\frac{48 M}{\Lambda }-\frac{27 \xi ^3 \mathfrak{Q}^3}{\Lambda ^3}+\frac{36 \xi  \mathfrak{Q}}{\Lambda ^2}}{4 \sqrt{\frac{\sqrt[3]{\beta }}{3 \sqrt[3]{2} \Lambda }-\frac{2}{\Lambda }+\frac{3 \sqrt[3]{2} (6 M \xi  \mathfrak{Q}+4 \Lambda  \mathfrak{Q}+1)}{\sqrt[3]{\beta } \Lambda }+\frac{9 \xi ^2 \mathfrak{Q}^2}{4 \Lambda ^2}}} \right. \\
& \left. -\frac{3 \sqrt[3]{2} (6 M \xi  \mathfrak{Q}+4 \Lambda  \mathfrak{Q}+1)}{\sqrt[3]{\beta } \Lambda }+\frac{9 \xi ^2 \mathfrak{Q}^2}{2 \Lambda ^2}\right]^{1/2}.
\end{split}
\fe
where 
\ie
\begin{split}
\nonumber
\beta & = + 972 \Lambda  M^2 + +486 M \xi  \mathfrak{Q}+729 \xi ^2 \mathfrak{Q}^3-648 \Lambda  \mathfrak{Q}+54 \\
& +\sqrt{\left(972 \Lambda  M^2 + 486 M \xi  \mathfrak{Q}+729 \xi ^2 \mathfrak{Q}^3-648 \Lambda  \mathfrak{Q}+54\right)^2-4 (54 M \xi  \mathfrak{Q}+36 \Lambda  \mathfrak{Q}+9)^3}.
\end{split}
\fe
Although four solutions arise, only one of them is physically relevant (because it takes real, positive--definite values) for the configuration considered here, denoted as \( r_{h} \). From this point onward, we shall use this physical horizon in the subsequent calculations presented in the following sections.

\begin{figure}
    \centering
     \includegraphics[scale=0.51]{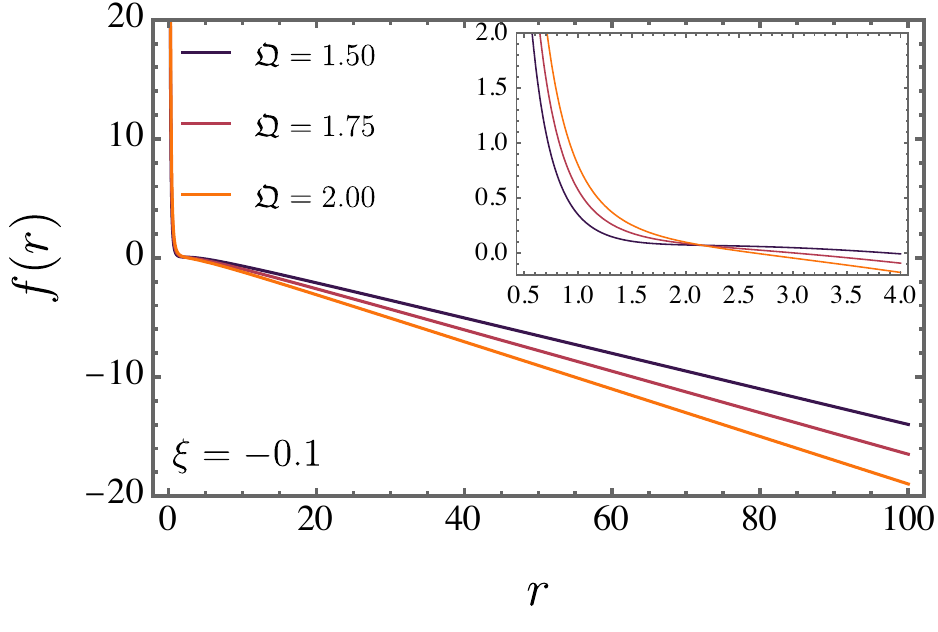}
     \includegraphics[scale=0.51]{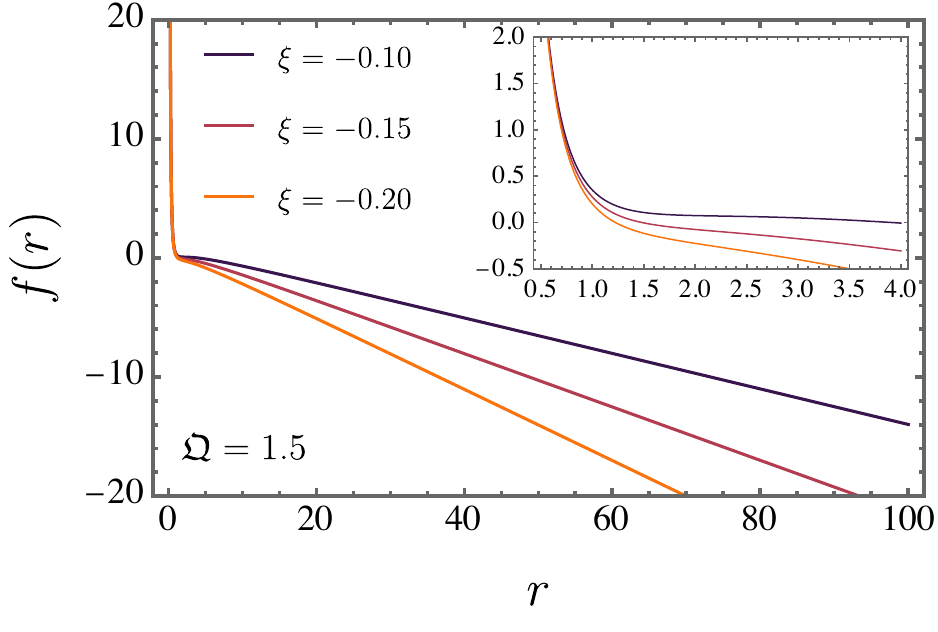}
    \caption{The representation of \( f(r) \) is analyzed for different values of the effective charge \( \mathfrak{Q} \) and the coupling parameter \( \xi \). Here, $\Lambda =  10^{-5}$.}
    \label{metricfuntionf}
\end{figure}

\begin{figure}
    \centering
     \includegraphics[scale=0.51]{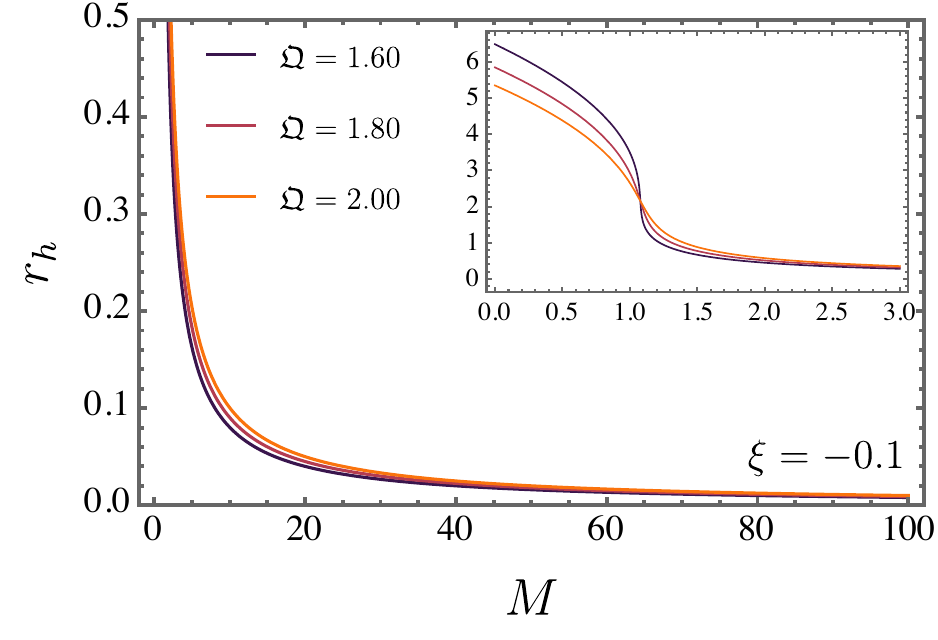}
     \includegraphics[scale=0.51]{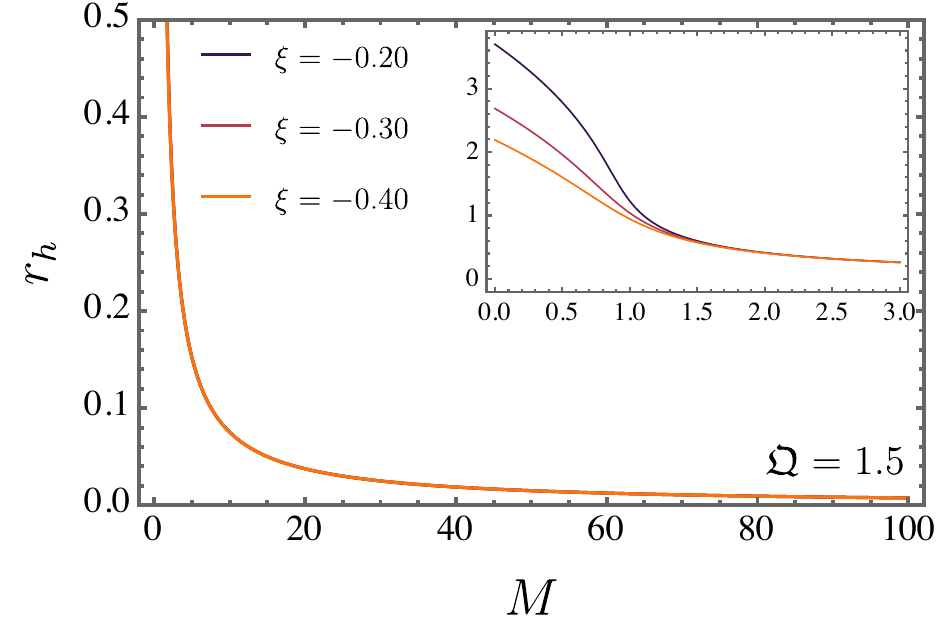}
    \caption{The representation of the event horizon\( r_{h} \) is analyzed for different values of the effective \( \mathfrak{Q} \) and the coupling parameter \( \xi \). Here, $\Lambda =  10^{-5}$.}
    \label{Mfunctionshorizons}
\end{figure}

Let us begin by analyzing the behavior of \( f(r) \) by varying the parameters \(\mathfrak{Q}\) and \(\xi\). This behavior is illustrated in Fig. \ref{metricfuntionf}, where both small and large values of \(r\) are considered. The magnitude of the cosmological constant used in this context is \(\Lambda =  10^{-5}\). Additionally, Fig. \ref{Mfunctionshorizons} presents the event horizon \( r_{h} \) as a function of the mass \( M \). In this plot, several values of \(\mathfrak{Q}\) and \(\xi\) are considered, with the cosmological constant fixed at \(\Lambda = 10^{-5}\). To provide a more detailed quantitative analysis of the event horizon, we present the results in Tab. \ref{quantitativehorizon}. Overall, we observe that a decrease in \(\xi\) combined with an increase in \(\mathfrak{Q}\) (for $\Lambda = 10^{-5}$) results in a reduction in the magnitude of the event horizon \(r_{h}\).

\begin{table}[!h]
\begin{center}
\begin{tabular}{c c c || c c c} 
 \hline\hline\hline
 $\xi$ & $\mathfrak{Q}$ &  $r_{h}$ & $\xi$ & $\mathfrak{Q}$ &  $r_{h}$ \\ [0.2ex] 
 \hline
 -0.1  & 1.5 & 3.91227 & -0.1  & 2.0 & 2.65081  \\ 

 -0.2  & 1.5 & 1.22406 & -0.1  & 3.0 & 2.31834  \\
 
 -0.3  & 1.5 & 1.03667 & -0.1  & 4.0 & 2.25177  \\
 
 -0.4  & 1.5 & 0.94306 & -0.1  & 4.5 & 2.23532  \\
 
 -0.5  & 1.5 & 0.88166 & -0.1  & 5.0 & 2.22362  \\ 
 [0.2ex] 
 \hline \hline \hline
\end{tabular}
\caption{\label{quantitativehorizon} A quantitative analysis of the event horizon \( r_{h} \) is performed for various values of \(\xi\) and \(\mathfrak{Q}\), while keeping the cosmological constant fixed at \(\Lambda = 10^{-5}\) and $M = 1.0$.}
\end{center}
\end{table}

Another key aspect of the theory under consideration is the Ricci scalar, which is defined as $R \equiv g_{\mu\nu}R^{\mu\nu}$,
\ie
R = -4 \Lambda -\frac{6 \xi  \mathfrak{Q}}{r}.
\fe

\begin{figure}
    \centering
     \includegraphics[scale=0.622]{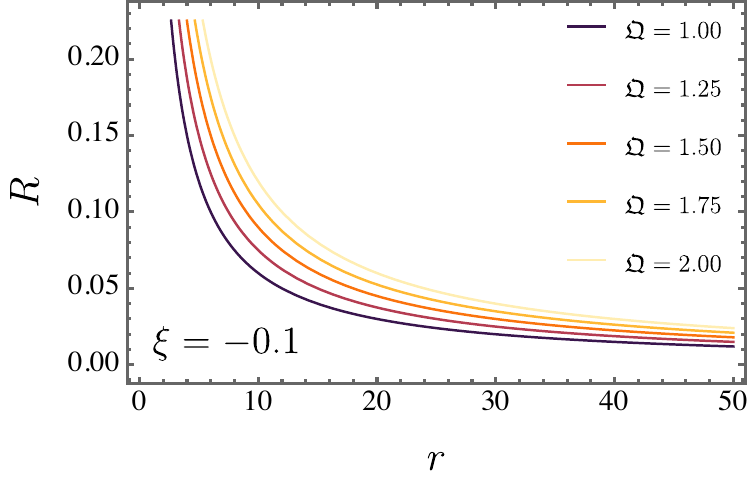}
     \includegraphics[scale=0.61]{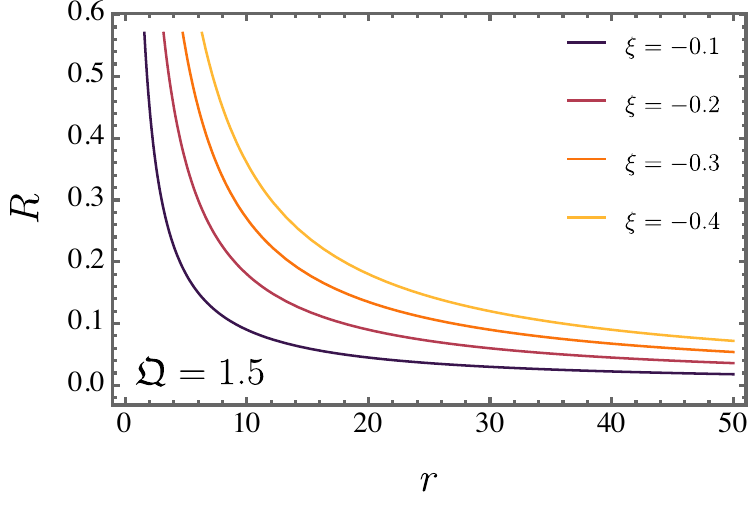}
    \caption{The Ricci scalar $R$ is depicted for various values of \(\xi\) and \(\mathfrak{Q}\), while keeping the cosmological constant fixed at \(\Lambda = 10^{-5}\).}
    \label{Ricci}
\end{figure}

In Fig. \ref{Ricci}, the behavior of the Ricci scalar is illustrated, revealing a clear singularity as \( r \to 0 \). Additionally, the Kretschmann scalar is another important quantity to examine for identifying potential singularities in the spacetime structure. Thereby,
\ie
K = \frac{8 \Lambda ^2}{3}+\frac{48 M^2}{r^6}-\frac{96 M \mathfrak{Q}}{r^7}+\frac{56 \mathfrak{Q}^2}{r^8}-\frac{8 \xi  \mathfrak{Q}^2}{r^5}+\frac{8 \xi ^2 \mathfrak{Q}^2}{r^2}+\frac{8 \Lambda  \xi  \mathfrak{Q}}{r}.
\fe

\begin{figure}
    \centering
     \includegraphics[scale=0.622]{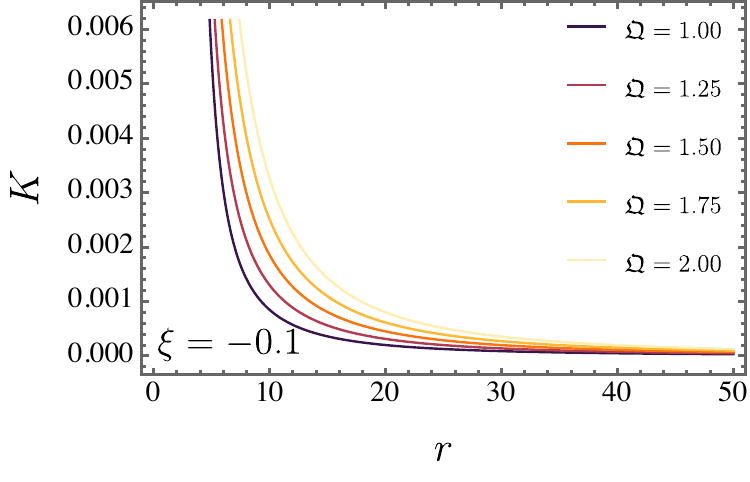}
     \includegraphics[scale=0.61]{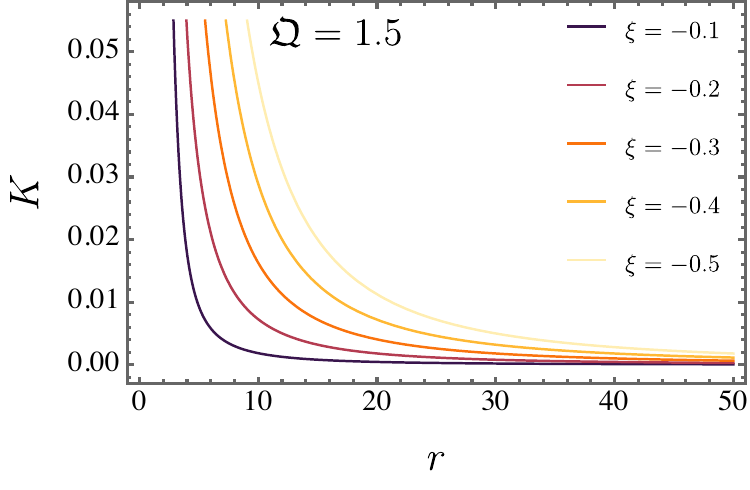}
    \caption{The Kretschmann scalar $K$ is shown for different values of \(\xi\) and \(\mathfrak{Q}\), while keeping the cosmological constant fixed at \(\Lambda = 10^{-5}\).}
    \label{Kretschmann}
\end{figure}

The expression above clearly reveals the presence of a physical singularity as \( r \to 0 \). To better illustrate how the Kretschmann scalar \( \mathfrak{K} \) behaves under varying \( \xi \) and \( \mathfrak{Q} \), we present Fig. \ref{Kretschmann}. At this stage, it is important to underscore the role of the non--zero Christoffel symbols \( \Gamma\indices{^\mu_{\alpha\beta}} \), which are fundamental in computing geodesics within the context of this theory. These symbols are then 
\ie
\begin{split}
\nonumber
& \Gamma\indices{^1_{00}} = \frac{\left(-6 M r+\Lambda  r^4+3 \xi  r^3 \mathfrak{Q}+3 r^2+3 \mathfrak{Q}\right) \left(6 M r+2 \Lambda  r^4+3 \mathfrak{Q} \left(\xi  r^3-2\right)\right)}{18 r^5},\\
& \Gamma\indices{^1_{11}} = \frac{6 M r+2 \Lambda  r^4+3 \xi  r^3 \mathfrak{Q}-6 \mathfrak{Q}}{12 M r^2-2 \Lambda  r^5-6 \xi  r^4 \mathfrak{Q}-6 r^3-6 r \mathfrak{Q}},\\
& \Gamma\indices{^1_{22}} = -r \left(-\frac{2 M}{r}+\frac{\Lambda  r^2}{3}+\frac{\mathfrak{Q}}{r^2}+\xi  r \mathfrak{Q}+1\right),\\
& \Gamma\indices{^1_{33}} = -r \sin ^2(\theta ) \left(-\frac{2 M}{r}+\frac{\Lambda  r^2}{3}+\frac{\mathfrak{Q}}{r^2}+\xi  r \mathfrak{Q}+1\right),  \\
& \Gamma\indices{^2_{21}} = 1/r,\\
& \Gamma\indices{^2_{33}} = \sin (\theta ) (-\cos (\theta )),\\
& \Gamma\indices{^3_{13}} = 1/r,\\
& \Gamma\indices{^3_{23}} = \cot (\theta ), \\
& \Gamma\indices{^3_{31}} = 1/r,\\ 
& \Gamma\indices{^3_{32}} =  \cot (\theta ),\\
& \Gamma\indices{^0_{10}} = \frac{6 M r+2 \Lambda  r^4+3 \mathfrak{Q} \left(\xi  r^3-2\right)}{2 r \left(-6 M r+\Lambda  r^4+3 \xi  r^3 \mathfrak{Q}+3 r^2+3 \mathfrak{Q}\right)}.
\end{split}
\fe
Thus, with the above expressions, the light trajectories can be precisely examined, as demonstrated in the next section.

\section{The trajectory of light}

The examination of light trajectories in a static solution for a nonlinear electromagnetic extension of the Kerr–Newman black hole provides an insightful comparison to the rotating case \cite{galindo2024nonlinear}. This analysis sheds light on how altered field dynamics, influenced by the presence of a cosmological constant, affect the structure of spacetime and its observable characteristics. In particular, photon spheres play a crucial role in determining the stability of light orbits and are key to understanding gravitational lensing, while black hole shadows offer a direct link to observable images, allowing for potential tests of theoretical predictions with astronomical observations. Additionally, analyzing geodesic motion uncovers critical aspects of light propagation within this spacetime. These features will be explored in detail in the following subsections.

\subsection{Critical orbits and shadow radii}

To move forward, we present a general form for the metric $g_{\mu\nu}$, which will serve as the basis for the subsequent analysis
\ie
\mathrm{d}s^{2} = g_{\mu \nu}\mathrm{d}x^\mu \mathrm{d}x^\nu  =  - \mathfrak{A}(r)\mathrm{d}t^2 + \mathfrak{B}(r)\mathrm{d}{r^2} + \mathfrak{C}(r)\mathrm{d}\theta ^2 + \mathfrak{D}(r)\,{{\mathop{\rm \sin}\nolimits} ^2}\theta \mathrm{d}\varphi ^2.
\fe
Here, \(\mathfrak{A}(r)\), \(\mathfrak{B}(r)\), \(\mathfrak{C}(r)\), and \(\mathfrak{D} (r)\) represent the respective components of the metric tensor. Moving forward, we apply the Lagrangian method in the following manner:
\ie
\mathcal{L} = \frac{1}{2}{g_{\mu \nu }}{{\dot x}^\mu }{{\dot x}^\nu },
\fe
in order that
\begin{equation}
\mathcal{L} = \frac{1}{2}[ - \mathfrak{A}(r){{\dot t}^2} + \mathfrak{B}(r){{\dot r}^2} + \mathfrak{C}(r){{\dot \theta }^2} + \mathfrak{D}(r){{\mathop{\rm \sin}\nolimits} ^2}\, \theta {{\dot \varphi }^2}].
\end{equation}

By applying the Euler--Lagrange equation and confining the motion to the equatorial plane (\(\theta = \frac{\pi}{2}\)), we derive two constants of motion: the energy \(E\) and the angular momentum \(L\), which are given by:
\begin{equation}\label{asdcosnstdsant}
E = \mathfrak{A}(r)\dot t \quad\mathrm{and}\quad L = \mathfrak{D}(r)\dot \varphi,
\end{equation}
and taking into account light particle modes, we obtain
\begin{equation}\label{lwiegrhtt}
- \mathfrak{A}(r){{\dot t}^2} + \mathfrak{B}(r){{\dot r}^2} + \mathfrak{D}(r){{\dot \varphi }^2} = 0.
\end{equation}

In this sense, after doing some algebraic manipulations for the sake of substituting Eq. (\ref{asdcosnstdsant}) in Eq. (\ref{lwiegrhtt}), it reads
\begin{equation}
\frac{{{{\dot r}^2}}}{{{{\dot \varphi }^2}}} = {\left(\frac{{\mathrm{d}r}}{{\mathrm{d}\varphi }}\right)^2} = \frac{{\mathfrak{D}(r)}}{{\mathfrak{B}(r)}}\left(\frac{{\mathfrak{D}(r)}}{{\mathfrak{A}(r)}}\frac{{{E^2}}}{{{L^2}}} - 1\right).
\end{equation}

In addition, we must highlight that
\ie
\frac{\mathrm{d}r}{\mathrm{d}\lambda} = \frac{\mathrm{d}r}{\mathrm{d}\varphi} \frac{\mathrm{d}\varphi}{\mathrm{d}\lambda}  = \frac{\mathrm{d}r}{\mathrm{d}\varphi}\frac{L}{\mathfrak{D}(r)}, 
\fe
where
\ie
\Dot{r}^{2} = \left( \frac{\mathrm{d}r}{\mathrm{d}\lambda} \right)^{2} =\left( \frac{\mathrm{d}r}{\mathrm{d}\varphi} \right)^{2} \frac{L^{2}}{\mathfrak{D}(r)^{2}}.
\fe

Up to this point, we have provided a general method for determining the critical orbits (photon sphere) in a generic spherically symmetric spacetime. Now, we will apply this basis to our specific case, yielding: $\mathfrak{A}(r) = -\frac{2 M}{r}+\frac{\Lambda  r^2}{3}+\frac{\mathfrak{Q} \left(\xi  r^3+1\right)}{r^2}+1$, $\mathfrak{B}(r) =  \left( -\frac{2 M}{r}+\frac{\Lambda  r^2}{3}+\frac{\mathfrak{Q} \left(\xi  r^3+1\right)}{r^2}+1\right)^{-1}$, $\mathfrak{C}(r) = r^{2}$ and $\mathfrak{D}(r) = r^{2}\sin^{2}\theta$. Thereby,
\ie
\Dot{r}^{2} = E^{2} + \mathfrak{V}(r,\xi,Q),
\fe
in which $\mathfrak{V}(r,\xi,Q)$ is
\ie
\mathfrak{V}(r,\xi,Q) = \frac{L^2 \left(-\frac{2 M}{r}+\frac{\Lambda  r^2}{3}+\frac{\mathfrak{Q} \left(\xi  r^3+1\right)}{r^2}+1\right)}{r^2}.
\fe

To locate the position of the photon sphere, we solve the equation \(\mathrm{d}\mathfrak{V}/\mathrm{d}r = 0\). Notably, this equation produces three distinct roots; however, only two of these, \(r_{ph1}\) and \(r_{ph2}\), represent physical solutions, as detailed below:
\ie
\begin{split}
& r_{ph1} =  -\frac{2}{3 \xi  \mathfrak{Q}} \\
& +\frac{\sqrt[3]{2} \Bar{\gamma} \left(9 M \xi  \mathfrak{Q}+2\right)}{3 \xi  \mathfrak{Q}\sqrt[3]{108 M \xi  \mathfrak{Q}+4 \sqrt{\left(27 \xi  \mathfrak{Q} \left(M+\xi  \mathfrak{Q}^{2}\right)+4\right)^2-2 \left(9 M \xi  \mathfrak{Q}+2\right)^3}+108 \xi ^2 \mathfrak{Q}^3+16}} \\
& \frac{\gamma \sqrt[3]{108 M \xi  \mathfrak{Q}+4 \sqrt{\left(27 \xi  \mathfrak{Q} \left(M+\xi  \mathfrak{Q}^2\right)+4\right)^2-2 \left(9 M \xi  \mathfrak{Q}+2\right)^3}+108 \xi ^2 \mathfrak{Q}^{3}+16}}{6 \sqrt[3]{2} \xi  \mathfrak{Q}},
\end{split}
\fe
and
\ie
\begin{split}
& r_{ph2} = -\frac{2}{3 \xi  \mathfrak{Q}} \\
& +\frac{\sqrt[3]{2} \gamma }{3 \xi  \mathfrak{Q} \sqrt[3]{108 M \xi  \mathfrak{Q}+4 \sqrt{\left(27 \xi  \mathfrak{Q} \left(M+\xi  \mathfrak{Q}^2\right)+4\right)^2-2 \left(9 M \xi  \mathfrak{Q}+2\right)^3}+108 \xi ^2 \mathfrak{Q}^3+16}} \\
& + \frac{\Bar{\gamma} \sqrt[3]{108 M \xi  \mathfrak{Q}+4 \sqrt{\left(27 \xi  \mathfrak{Q} \left(M+\xi  \mathfrak{Q}^2\right)+4\right)^2-2 \left(9 M \xi  \mathfrak{Q}+2\right)^3}+108 \xi ^2 Q^6+16}}{6 \sqrt[3]{2} \xi  \mathfrak{Q}},
\end{split}
\fe
in which \(\gamma \equiv 1 - i \sqrt{3}\) and \(\Bar{\gamma} \equiv 1 + i \sqrt{3}\). it is important to highlight that the cosmological constant \(\Lambda\) does not impact the photon spheres. Consequently, they are identical to those recently analyzed in the literature \cite{AraujoFilho:2024xhm}, where a comprehensive discussion of these characteristics can also be found.

\begin{figure}
    \centering
     \includegraphics[scale=0.63]{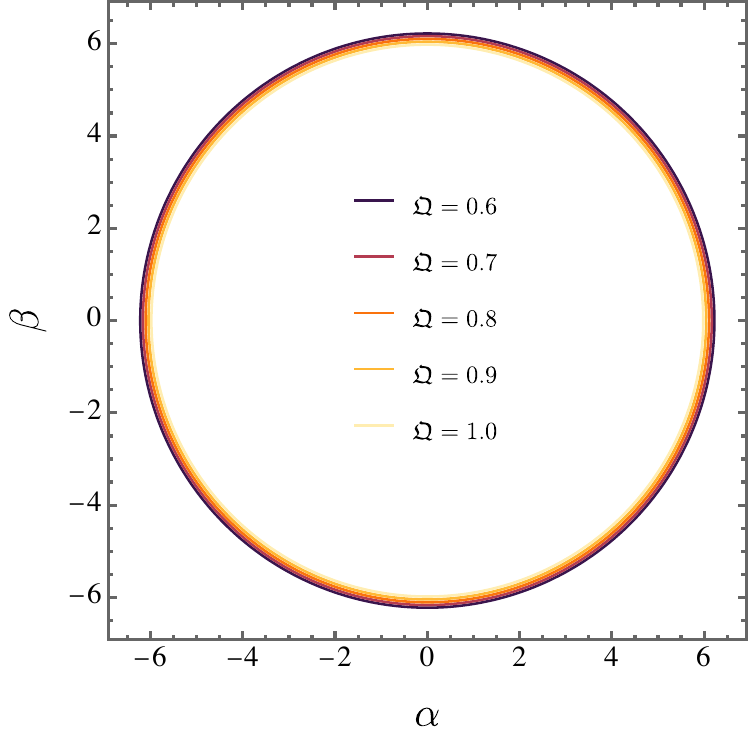}
     \includegraphics[scale=0.63]{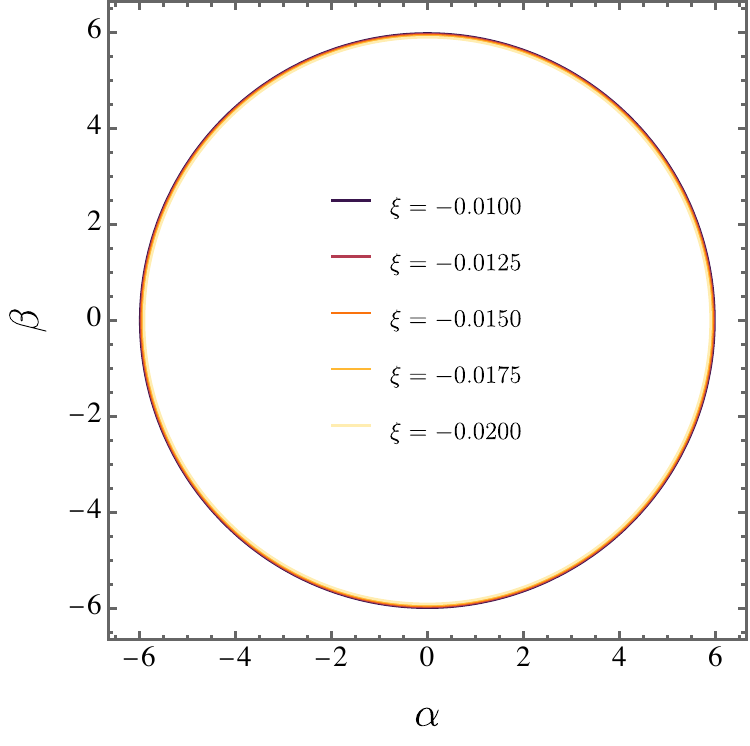}
    \caption{The shadow plots are illustrated for a range of \(\mathfrak{Q}\) and \(\xi\) values, while keeping the cosmological constant fixed at \(\Lambda = 10^{-5}\).}
    \label{shadplots}
\end{figure}

With all this in place, the shadow radii can be accurately expressed as shown below
\ie
\mathcal{R} = r_{ph2}  \sqrt{\frac{-\frac{2 M}{r_{o}}+\frac{\Lambda  r_{o}^2}{3}+\frac{\mathfrak{Q}}{r_{o}^2}+\xi  r_{o} \mathfrak{Q}+1}{\frac{\Lambda  r_{ph2} ^2}{3}-\frac{2 M}{r_{ph2} }+\mathfrak{Q} \left(\xi  r_{ph2} +\frac{1}{r_{ph2} ^2}\right)+1}},
\fe
where $r_{o}$ is the radial coordinate for a distant observer. Using the standard method found in the literature, we present the results as parametric plots of the celestial coordinates \(\alpha\) and \(\beta\) \cite{afrin2024testing,cel1,cel2,cel3,cel4}. Fig. \ref{shadplots} shows the boundaries of the black hole shadow for different values of \(\mathfrak{Q}\) and \(\xi\). In the left panel, the shadow contours are displayed for varying \(\mathfrak{Q}\) while keeping \(\xi = -0.1\) and \(\Lambda = 10^{-5}\) fixed. As \(\mathfrak{Q}\) increases, the shadow silhouette contracts. In the right panel, shadow profiles are displayed for various \(\xi\) values (with \(\mathfrak{Q} = 0.5\) and \(\Lambda = 10^{-5}\)), illustrating that a reduction in \(\xi\) results in a smaller shadow radius under the influence of the cosmological constant \(\Lambda\).

\subsection{Geodesics}

In this section, we focus on analyzing the geodesics. As highlighted in the previous section, calculating the Christoffel symbols is essential for this process. Thus, we proceed by writing 
\ie
\frac{\mathrm{d}^{2}x^{\mu}}{\mathrm{d}\tau^{2}} + \Gamma\indices{^\mu_\alpha_\beta}\frac{\mathrm{d}x^{\alpha}}{\mathrm{d}\tau}\frac{\mathrm{d}x^{\beta}}{\mathrm{d}\tau} = 0. \label{geogeo}
\fe

In this case, \( \tau \) serves as a general affine parameter. This formulation results in a set of four interdependent differential equations, each corresponding to the motion along a particular coordinate, as detailed below:
\ie
\frac{\mathrm{d} t^{\prime}}{\mathrm{d} \tau} = -\frac{r' t' \left(6 M r+2 \Lambda  r^4+3 \mathfrak{Q} \left(\xi  r^3-2\right)\right)}{r \left(-6 M r+\Lambda  r^4+3 \xi  r^3 \mathfrak{Q}+3 r^2+3 \mathfrak{Q}\right)},
\fe
\ie
\begin{split}
& \frac{\mathrm{d} r^{\prime}}{\mathrm{d} \tau} = \frac{\left(r'\right)^2 \left(6 M r+2 \Lambda  r^4+3 \mathfrak{Q} \left(\xi  r^3-2\right)\right)}{2 r \left(-6 M r+\Lambda  r^4+3 \xi  r^3 \mathfrak{Q}+3 r^2+3 \mathfrak{Q}\right)} \\
& -\frac{\left(t'\right)^2 \left(-6 M r+\Lambda  r^4+3 \xi  r^3 \mathfrak{Q}+3 r^2+3 \mathfrak{Q}\right) \left(6 M r+2 \Lambda  r^4+3 \mathfrak{Q} \left(\xi  r^3-2\right)\right)}{18 r^5} \\
& +r \left(\theta '\right)^2 \left(-\frac{2 M}{r}+\frac{\Lambda  r^2}{3}+\frac{\mathfrak{Q}}{r^2}+\xi  r \mathfrak{Q}+1\right)  +r \sin ^2(\theta ) \left(\varphi '\right)^2 \left(-\frac{2 M}{r}+\frac{\Lambda  r^2}{3}+\frac{\mathfrak{Q}}{r^2}+\xi  r \mathfrak{Q}+1\right), 
\end{split}
\fe
\ie
\frac{\mathrm{d} \theta^{\prime}}{\mathrm{d} \tau} = \sin (\theta ) \cos (\theta ) \left(\varphi '\right)^2-\frac{2 \theta ' r'}{r},
\fe
and, 
\ie
\frac{\mathrm{d} \varphi^{\prime}}{\mathrm{d} \tau} = -\frac{2 \varphi ' \left(r'+r \theta ' \cot (\theta )\right)}{r}.
\fe

\begin{figure}
    \centering
     \includegraphics[scale=0.5]
     {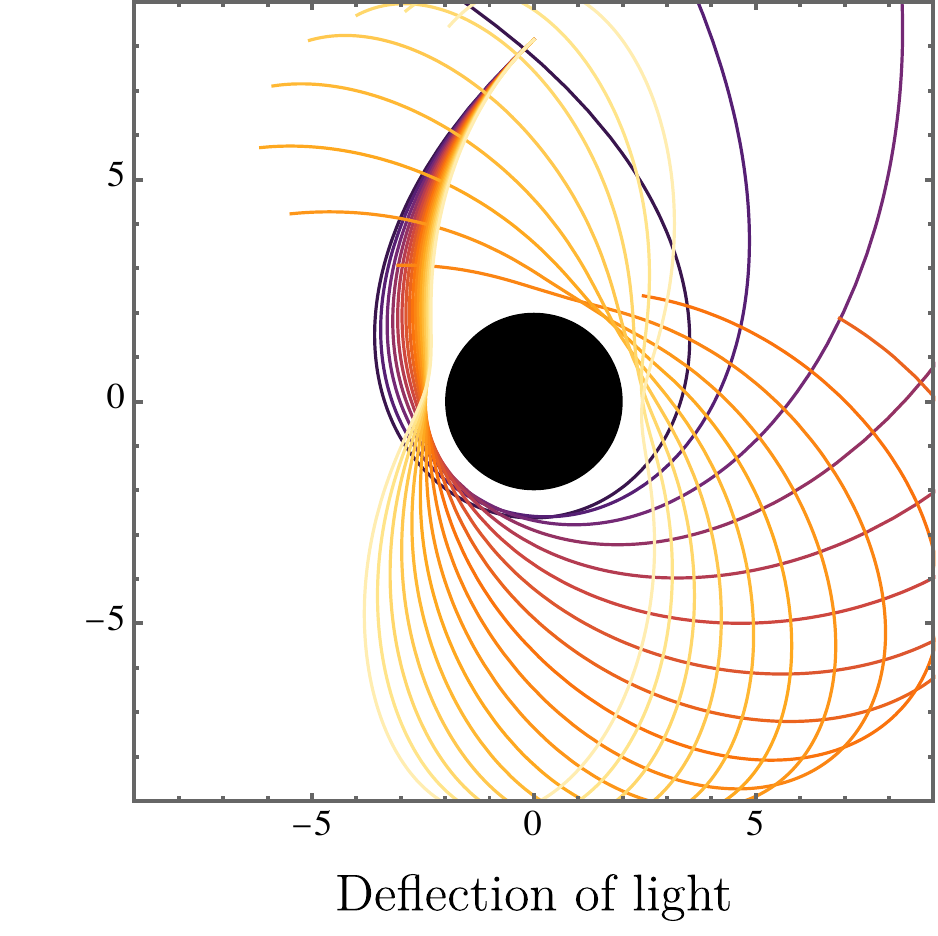}
     \includegraphics[scale=0.5]
     {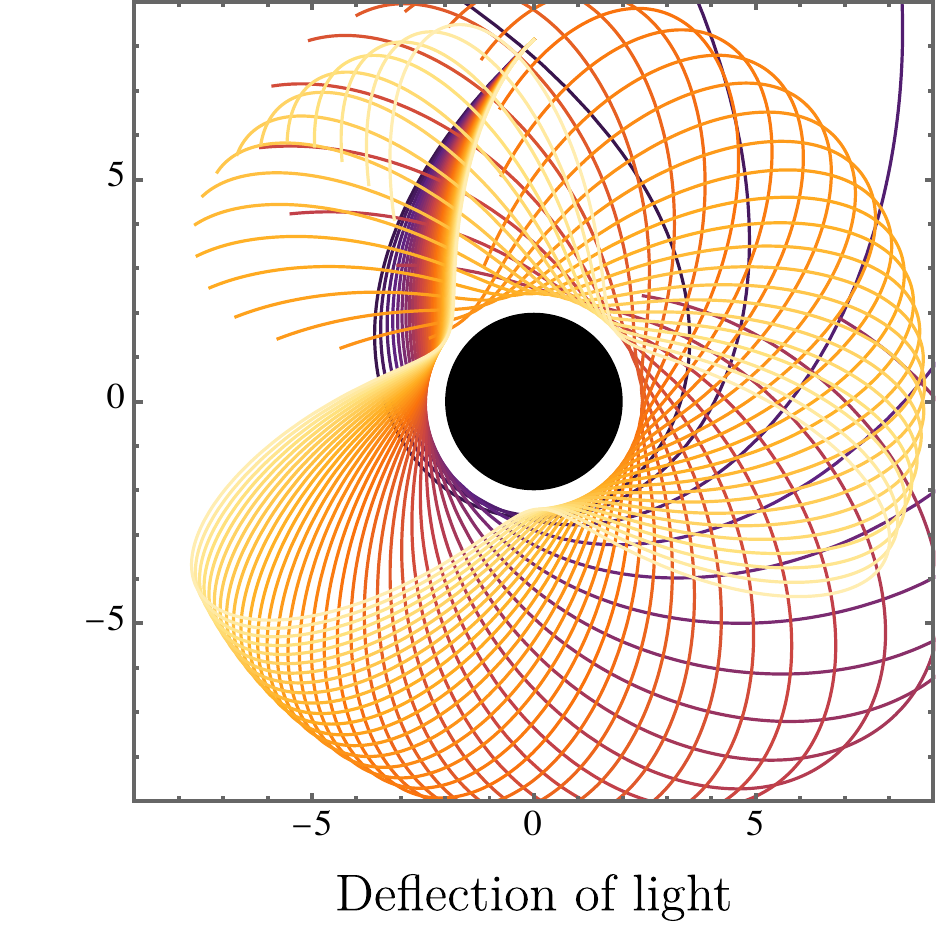}
     \includegraphics[scale=0.5]
     {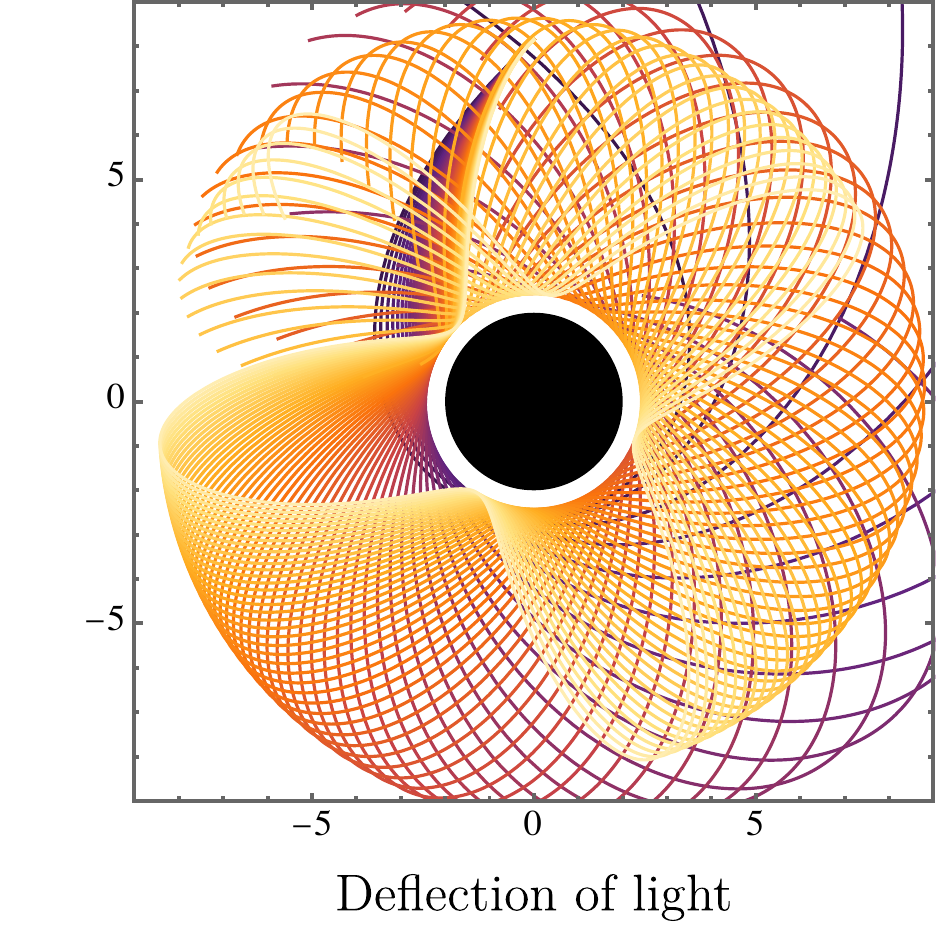}
     \includegraphics[scale=0.5]
     {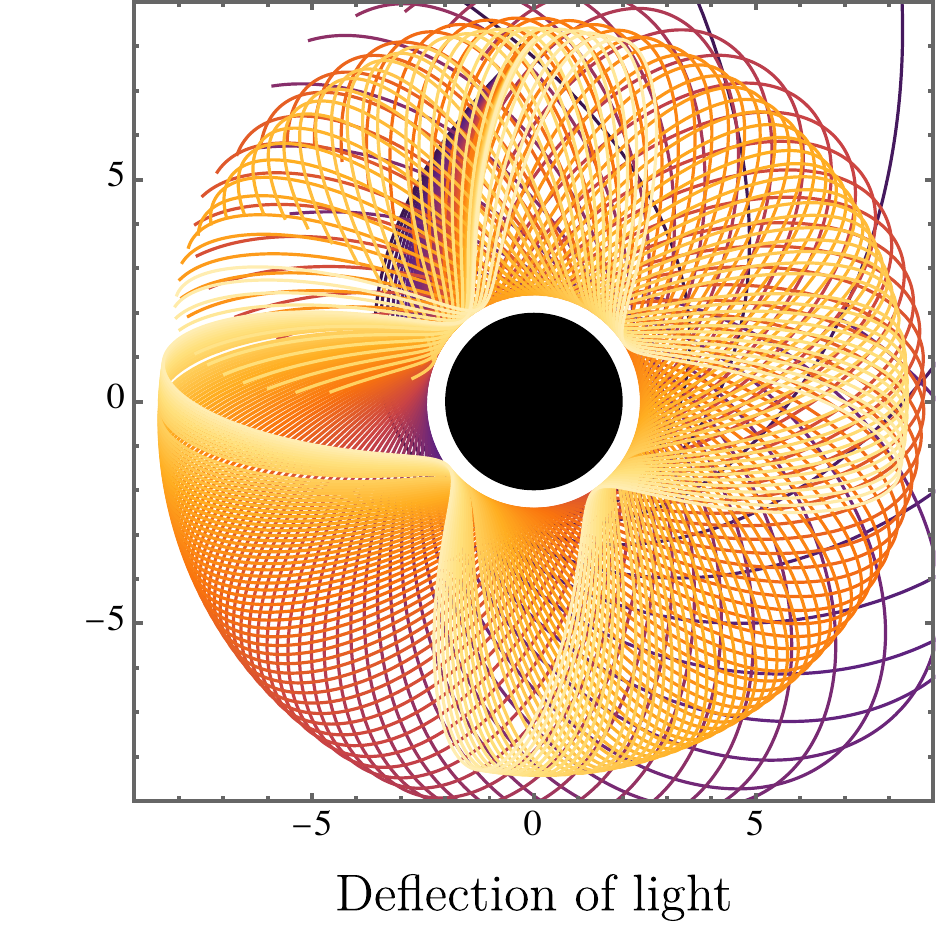}
    \caption{The ray tracing is shown for several values of $\mathfrak{Q}$ (keeping $\xi = -0.1$ and $\Lambda = 10^{-5}$).}
    \label{deflec}
\end{figure}

\begin{figure}
    \centering
     \includegraphics[scale=0.5]
     {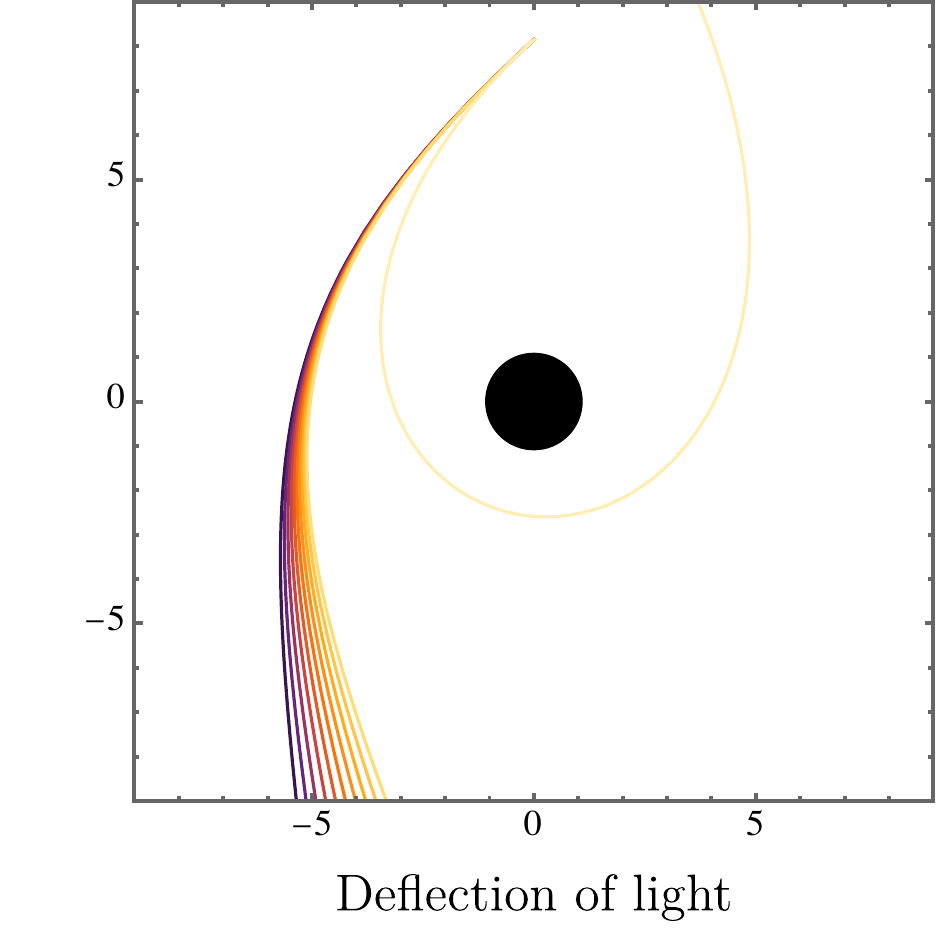}
     \includegraphics[scale=0.5]
     {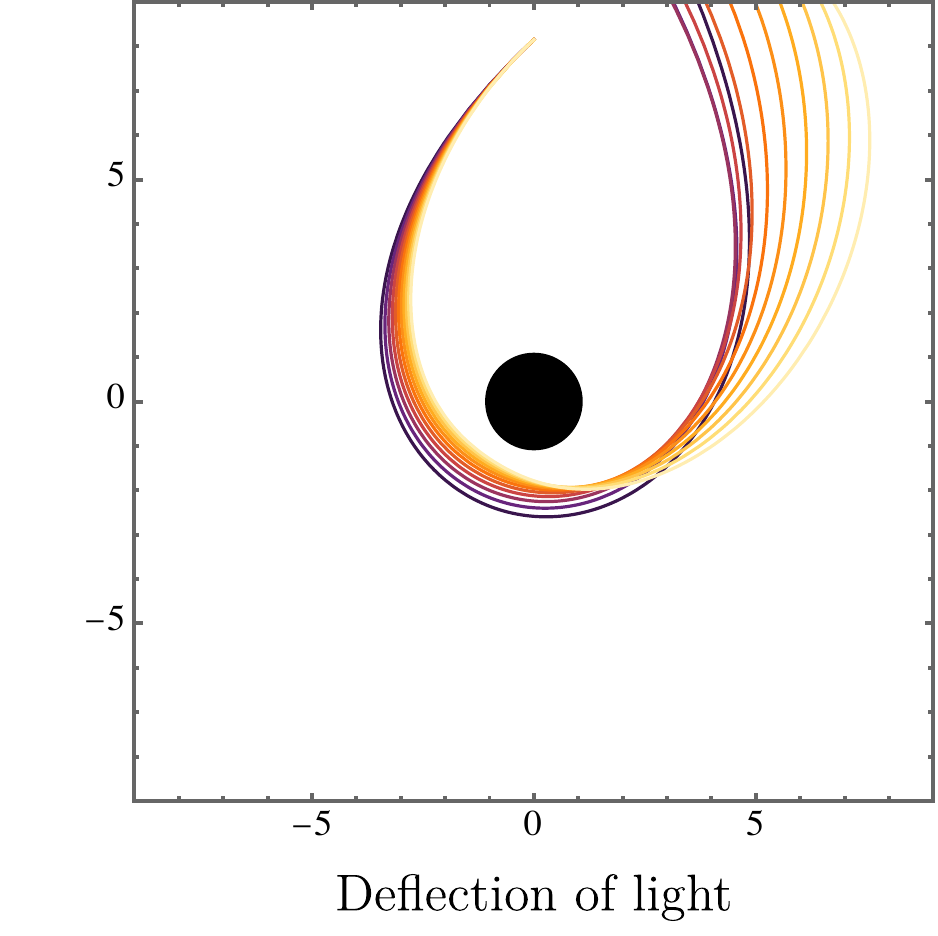}
     \includegraphics[scale=0.5]
     {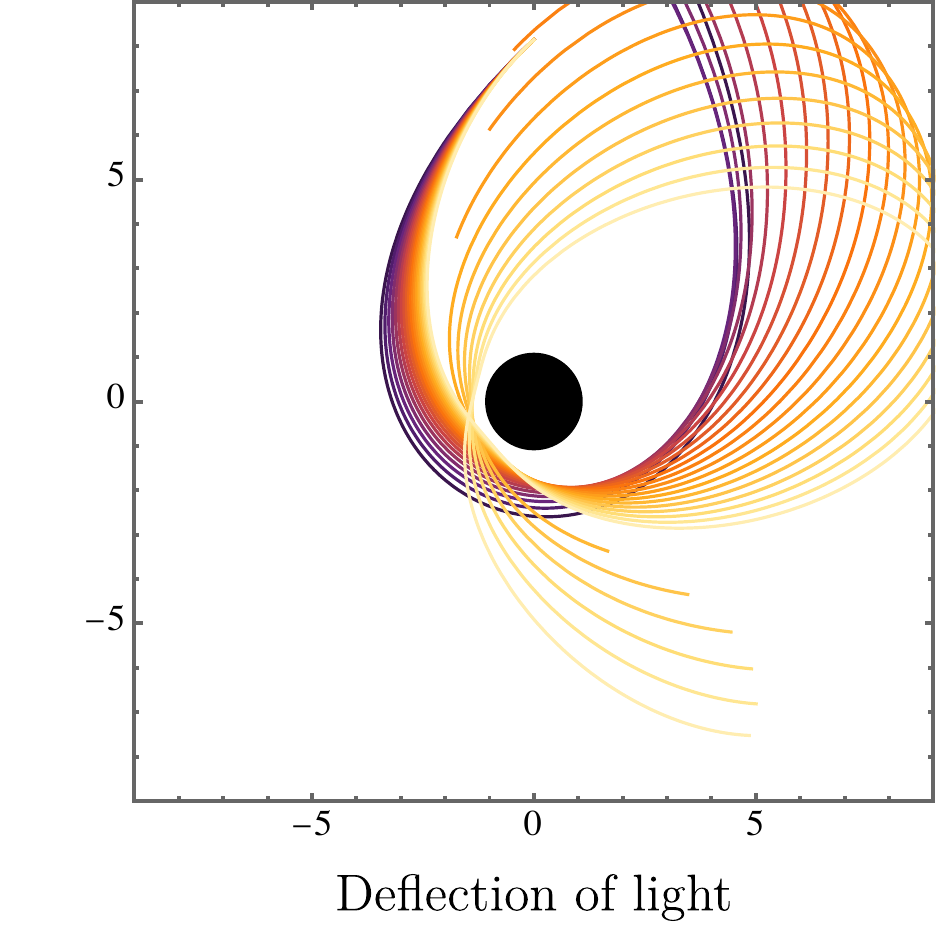}
     \includegraphics[scale=0.5]
     {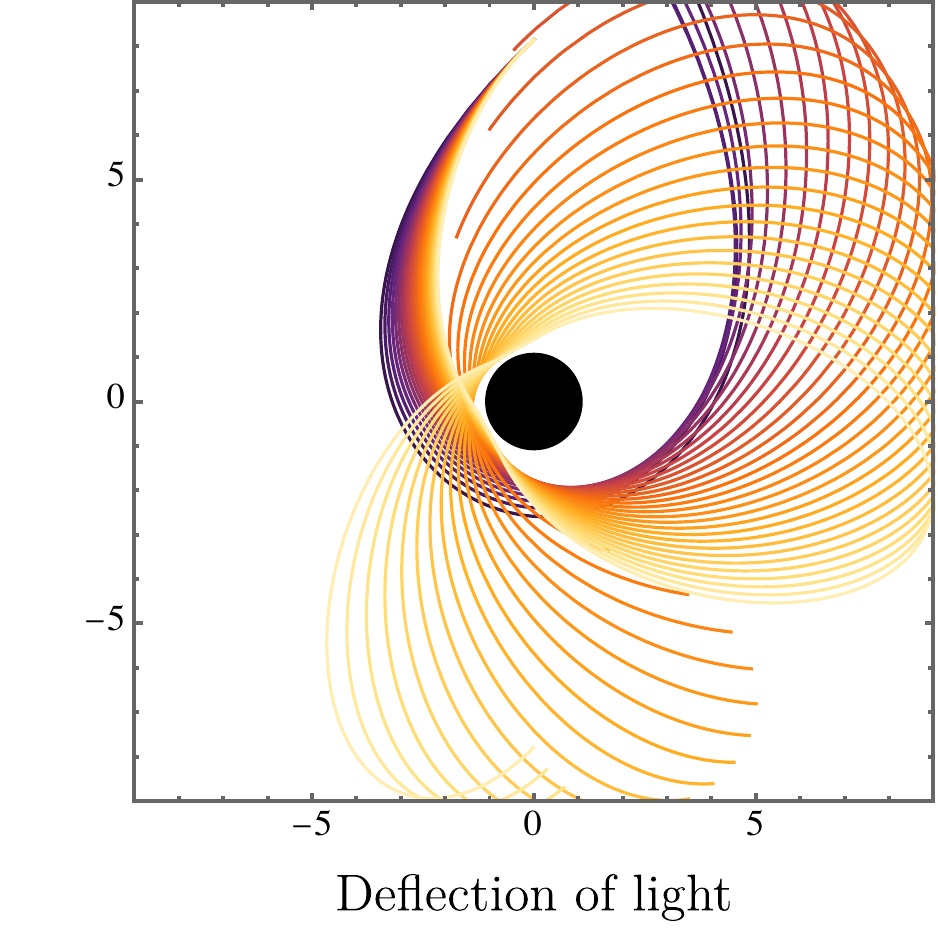}
    \caption{The ray tracing is shown for several values of $\xi$ (keeping $\mathfrak{Q} = 2.5$)}
    \label{deflec2}
\end{figure}

Using a numerical approach, we present Figs. \ref{deflec} and \ref{deflec2}. The first figure illustrates the light's trajectory as \(\mathfrak{Q}\) varies, while keeping \(\xi = -0.1\) constant. The second figure shows the light's path for different values of \(\xi\), with \(\mathfrak{Q}\) fixed at $2.5$. In both cases, the black disk represents the event horizon. Additionally, in Fig. \ref{deflec}, \(\mathfrak{Q}\) ranges from $2.1$ to $40$, whereas in Fig. \ref{deflec2}, \(\xi\) varies from $-0.01$ to $-0.4$.

\section{Thermal analysis}

Bardeen, Carter, and Hawking introduced a groundbreaking set of principles in 1970s, now known as the four laws of black hole mechanics, designed to mirror key concepts from thermodynamics \cite{bardeen1973four}. The zeroth law declares that surface gravity is uniformly distributed along the event horizon, much like the way temperature remains constant in a thermal equilibrium state \cite{page2005hawking}. The first law connects variations in a black hole’s mass  --- interpreted as energy --- with changes in its surface area, angular momentum, and electric charge, reflecting the thermodynamic relationship between internal energy, heat, and mechanical work \cite{carlip2014black}. The second law ensures that the total area of the event horizon never decreases, a principle that parallels the irreversible increase of entropy in isolated systems \cite{davies1978thermodynamics}. Finally, the third law establishes the impossibility of reducing a black hole's surface gravity to zero, similar to the thermodynamic restriction on reaching absolute zero temperature \cite{hawking1976black}.

Christodoulou’s work around the same period greatly expanded the comprehension of black hole dynamics by focusing on the irreversible processes governing them \cite{christodoulou1970reversible}. Simultaneously, Bekenstein introduced a groundbreaking idea by proposing that black holes possess entropy, which revolutionized the thermodynamic interpretation of these objects. His contributions linked a black hole’s entropy directly to the surface area of its event horizon \cite{1bekenstein2020black,2bekenstein1974generalized}. This key remark eventually culminated in the Bekenstein--Hawking entropy formula, forming a natural connection between the mechanics of black holes and the laws of thermodynamics.


\subsection{Hawking temperature}

In this part of the discussion, we focus on the essential features of Hawking temperature. As will be shown in the subsequent sections, this thermal property plays a pivotal role in characterizing the evaporation process, particularly as the black hole nears its final phase. To explore this, we express \cite{heidari2023gravitational,nascimento2024properties}
\ie
\label{hawkingtemmp}
T = \frac{1}{{4\pi \sqrt {{g_{tt}}{g_{rr}}} }}{\left. {\frac{{\mathrm{d}{g_{rr}}}}{{\mathrm{d}r}}} \right|_{r = {r_{h}}}} = \frac{\Lambda  r_{h}^4+2 \xi  r_{h}^3 \mathfrak{Q}+r_{h}^2-\mathfrak{Q}}{4 \pi  r_{h}^3}.
\fe

\begin{figure}
    \centering
     \includegraphics[scale=0.51]{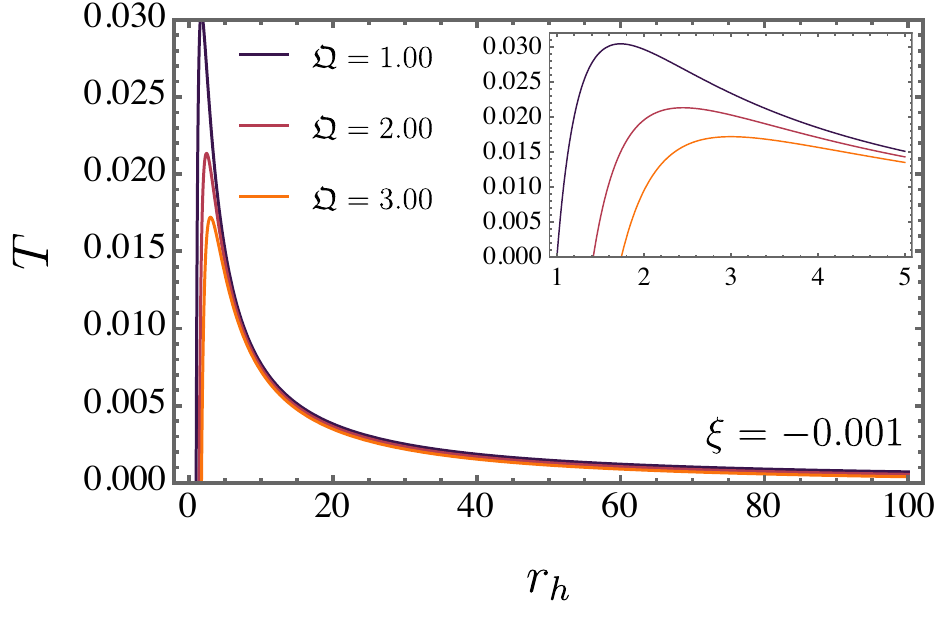}
     \includegraphics[scale=0.51]{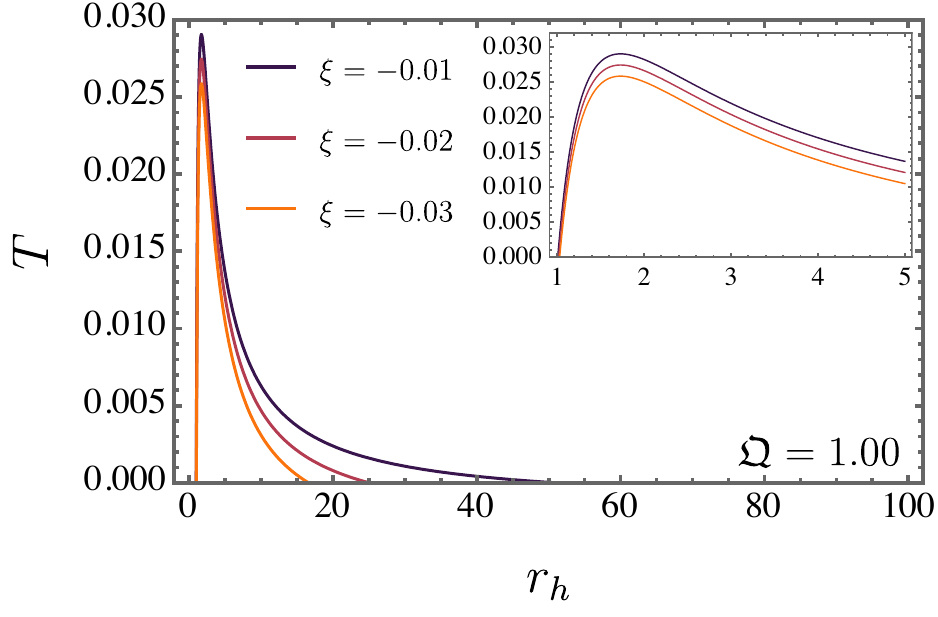}
    \caption{The Hawking temperature is shown for several values of $\xi$ and $Q$}
    \label{tempetarueT}
\end{figure}

To enhance the clarity of our findings, we provide Fig. \ref{tempetarueT}. The figure illustrates that an increase in \(\mathfrak{Q}\) leads to a further decrease in the value of \(T\). Additionally, as \(\xi\) decreases, the magnitude of the Hawking temperature is also reduced accordingly. All these features are computed from $\Lambda = 10^{-5}$.

\subsection{Heat capacity}

\begin{figure}
    \centering
     \includegraphics[scale=0.515]{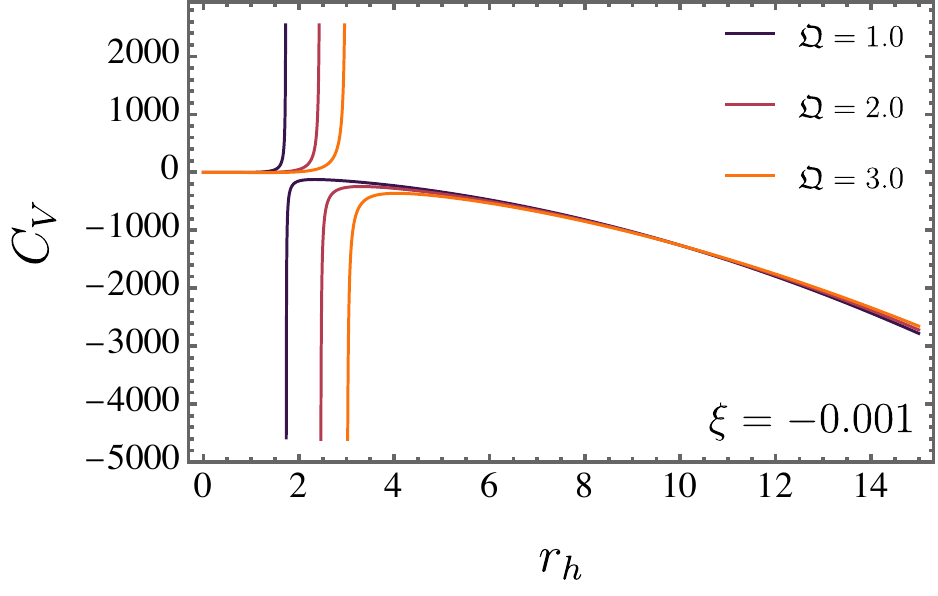}
     \includegraphics[scale=0.51]{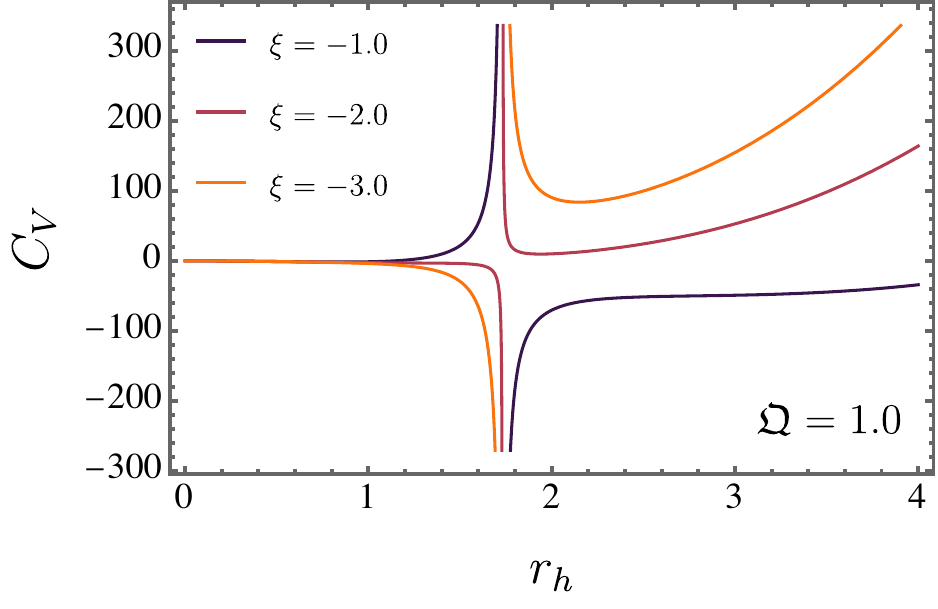}
    \caption{The heat capacity is exhibited for several values of $\xi$ and $\mathfrak{Q}$ }
    \label{heatcap}
\end{figure}

Complementing the results derived so far, we address here the behavior of the heat capacity $C_{V}$. In this manner, we write
\ie
C_{V} = T \frac{\partial S}{\partial T} =   \frac{4 \pi  r_{h}^2 \left(\Lambda  r_{h}^4+2 \xi  r_{h}^3 \mathfrak{Q}+r_{h}^2-\mathfrak{Q}\right)}{\Lambda  r_{h}^4-r_{h}^2+3 \mathfrak{Q}},
\fe
where the entropy $S$ is given by $S = \pi r_{h}^{2}$. In Fig. \ref{heatcap}, the behavior of heat capacity is illustrated for different values of \(\xi\) and \(\mathfrak{Q}\), with \(\Lambda\) set to \(10^{-5}\). The plot reveals both positive (stable) and negative (unstable) regions. For stable configurations, decreasing \(\xi\) leads to a reduction in the magnitude of \(C_{V}\). In contrast, unstable configurations exhibit an increase in \(C_{V}\) as \(\xi\) is reduced. Furthermore, when we consider the variation of \(\mathfrak{Q}\) for the heat capacity, the plots turn out to be shifted to the right.

\subsection{Gibbs free energy}

\begin{figure}
    \centering
     \includegraphics[scale=0.55]{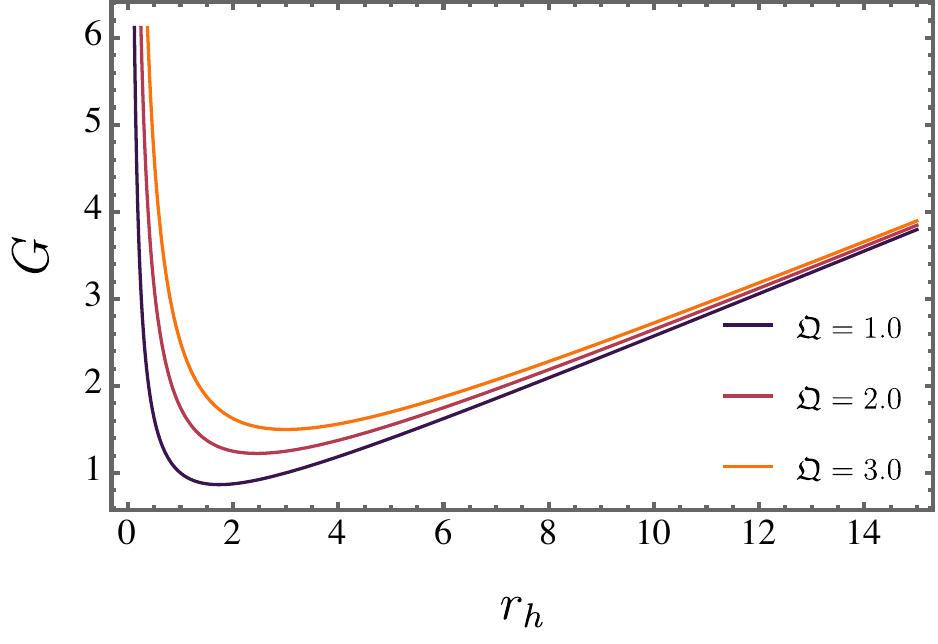}
    \caption{The Gibbs free energy is presented as a function of the event horizon radius \(r_h\) for various values of \(\mathfrak{Q}\).}
    \label{gibbsrh}
\end{figure}

\begin{figure}
    \centering
     \includegraphics[scale=0.51]{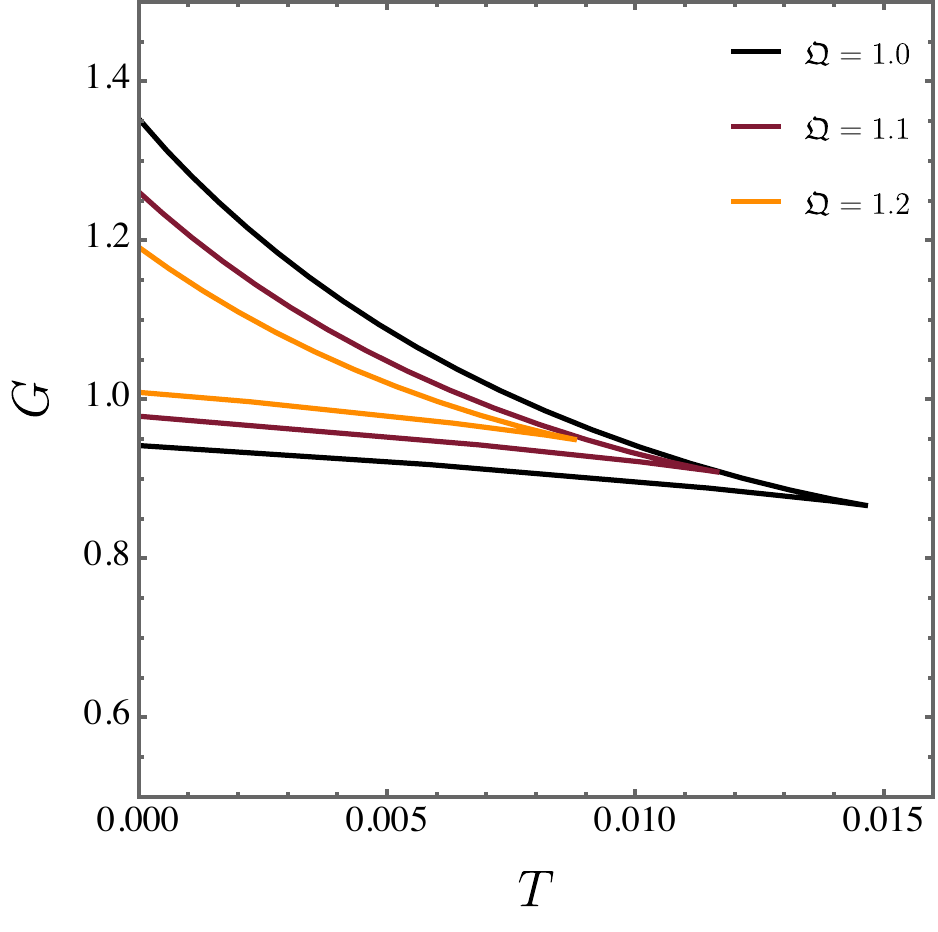}
    \caption{The Gibbs free energy is shown as a parametric plot with the Hawking temperature \(T\) for different values of \(\mathfrak{Q}\).}
    \label{gibbst}
\end{figure}

In order to provide a complementary analysis of our thermodynamic state quantities, let us calculate the Gibbs free energy. Thereby, it reads,
\ie
G = M - TS = \frac{-\Lambda  r_{h}^4+3 r_{h}^2+9 \mathfrak{Q}}{12 r_{h}},
\fe
where \( M = \frac{1}{6} r \left(\Lambda r^2 + 3 \xi r \mathfrak{Q} + 3\right) + \frac{\mathfrak{Q}}{2r} \), which is derived by setting \( f(r) = 0 \) and solving for the mass \( M \). Notably, the parameter \(\xi\), which represents the non--linear electrodynamics, does not influence this thermodynamic function shown above. Moving forward, we examine two distinct configurations in the following plots: in Fig. \ref{gibbsrh}, the Gibbs free energy is plotted as a function of the event horizon radius \(r_h\), and in Fig. \ref{gibbst}, \(G\) is displayed as a parametric plot with respect to the Hawking temperature. In general terms, an increase in \(\mathfrak{Q}\) leads to a corresponding rise in \(G\).

\section{Quantum radiation}

Before moving forward, it is essential to emphasize that the forthcoming calculations exclude any consideration of backreaction effects. Quantum tunneling enables particles within a black hole to escape by crossing the event horizon. This probability of tunneling can be obtained through methods discussed in Refs. \cite{angheben2005hawking,kerner2006tunnelling,kerner2008fermions}.


The phenomenon of black holes emitting radiation, comparable to black body radiation, arises from their inherent temperature. However, this radiation does not account for modifications introduced by greybody factors. The emitted spectrum is expected to consist of particles with varying spin, including fermions. Research led by Kerner and Mann \cite{o69}, along with subsequent studies \cite{o75,o72,o71,o74,o73,o70}, has shown that massless bosons and fermions are emitted at an identical temperature. Additionally, investigations into spin-$1$ bosons reveal that the Hawking temperature remains unaffected, even when higher--order quantum corrections in $\hbar$ beyond the semi--calssical approach are incorporated \cite{o77,o76}.

The dynamics of fermions are often framed in terms of the phase of their spinor wave function, governed by the Hamilton–Jacobi equation. Another representation for the action, as described in \cite{o83,o84,vanzo2011tunnelling}, takes the form $S_f = S_0 + (\text{spin corrections})$, where $ S_0$ is the classical action for scalar particles. The spin-dependent terms account for the interaction between the particle's spin and the spin connection of the underlying geometry, without causing singularities at the horizon. These corrections, being minor and primarily affecting spin precession, are disregarded in this analysis. Moreover, the influence of emitted particles' spin on the black hole’s angular momentum is negligible, particularly for non-rotating black holes with masses much larger than the Planck scale \cite{vanzo2011tunnelling}. The emission of particles with opposite spins occurs symmetrically on average, ensuring that the net angular momentum of the black hole remains effectively unaltered.

Here, we examine the tunneling process of fermionic particles traversing the event horizon of a specific black hole configuration. The emission rate is determined using a Schwarzschild--like coordinate system, which exhibits a singularity at the horizon. For alternative approaches based on generalized Painlevé–Gullstrand or Kruskal–Szekeres coordinates, readers are referred to the foundational analysis in \cite{o69}. The starting point for this investigation involves a general spacetime metric given by $
\mathrm{d}s^{2} = A(r) \mathrm{d}t^{2} + \frac{\mathrm{d}r^{2}}{B(r)} + C(r)\left(\mathrm{d}\theta^{2} + r^{2}\sin^{2}\theta\, \mathrm{d}\varphi^{2}\right)$.  
In the context of curved spacetime, the Dirac equation, which governs fermions, takes the form:
$
\left(\gamma^\mu \nabla_\mu + \frac{m}{\hbar}\right) \Psi(t,r,\theta,\varphi) = 0$
in which
$
\nabla_\mu = \partial_\mu + \frac{i}{2} {\Gamma^\alpha_{\;\mu}}^{\;\beta} \,\Sigma_{\alpha\beta}$ and $ 
\Sigma_{\alpha\beta} = \frac{i}{4} [\gamma_\alpha,  \gamma_\beta]$.
The \( \gamma^\mu \) matrices satisfy the defining properties of the Clifford algebra, which are encapsulated in the following relation:
$
\{\gamma_\alpha,\gamma_\beta\} = 2 g_{\alpha\beta} \mathbb{1}$, 
where \(\mathbb{1}\) represents the \(4 \times 4\) identity matrix. In this framework, the \(\gamma\) matrices are defined as:
\begin{eqnarray*}
 \gamma ^{t} &=&\frac{i}{\sqrt{A(r)}}\left( \begin{array}{cc}
\vec{1}& \vec{ 0} \\ 
\vec{ 0} & -\vec{ 1}%
\end{array}%
\right), \;\;
\gamma ^{r} =\sqrt{B(r)}\left( 
\begin{array}{cc}
\vec{0} &  \vec{\sigma}_{3} \\ 
 \vec{\sigma}_{3} & \vec{0}%
\end{array}%
\right), \\
\gamma ^{\theta } &=&\frac{1}{r}\left( 
\begin{array}{cc}
\vec{0} &  \vec{\sigma}_{1} \\ 
 \vec{\sigma}_{1} & \vec{0}%
\end{array}%
\right), \;\;
\gamma ^{\varphi } =\frac{1}{r\sin \theta }\left( 
\begin{array}{cc}
\vec{0} &  \vec{\sigma}_{2} \\ 
 \vec{\sigma}_{2} & \vec{0}%
\end{array}%
\right),
\end{eqnarray*}%
where \(\vec{\sigma}\) represents the Pauli matrices, which satisfy the standard commutation relation:
$\sigma_i  \sigma_j = \vec{1} \delta_{ij} + i \varepsilon_{ijk} \sigma_k, \,\, \text{in which}\,\, i,j,k =1,2,3.$ Also, notice that the matrix for \(\gamma^5\) is alternatively expressed as
\begin{equation*}
\gamma ^{5}=i\gamma ^{t}\gamma ^{r}\gamma ^{\theta }\gamma ^{\varphi }=i\sqrt{\frac{B(r)}{A(r)}}\frac{1}{r^{2}\sin \theta }\left( 
\begin{array}{cc}
\vec{ 0} & - \vec{ 1} \\ 
\vec{ 1} & \vec{ 0}%
\end{array}%
\right)\:.
\end{equation*}

To characterize a Dirac field with its spin directed upward along the positive \(r\)-axis, the chosen ansatz is \cite{vagnozzi2022horizon}:
\begin{equation}
\Psi_{\uparrow}(t,r,\theta ,\varphi ) = \left( \begin{array}{c}
H(t,r,\theta ,\varphi ) \\ 
0 \\ 
Y(t,r,\theta ,\varphi ) \\ 
0%
\end{array}%
\right) \exp \left[ \frac{i}{\hbar }\psi_{\uparrow}(t,r,\theta ,\varphi )\right]\;.
\label{spinupbh} 
\end{equation} 

Our analysis concentrates on the spin--up (\(\uparrow\)) configuration, with the understanding that the spin--down (\(\downarrow\)) case, aligned along the negative \(r\)-axis, can be treated in a comparable manner. Inserting the ansatz (\ref{spinupbh}) into the Dirac equation yields:
\ie
\begin{split}
-\left( \frac{i \,H}{\sqrt{A(r)}}\,\partial _{t} \psi_{\uparrow} + Y \sqrt{B(r)} \,\partial_{r} \psi_{\uparrow}\right) + H m &=0, \quad
-\frac{Y}{r}\left( \partial _{\theta }\psi_{\uparrow} +\frac{i}{\sin \theta } \, \partial _{\varphi }\psi_{\uparrow}\right) = 0, \\
\left( \frac{i \,Y}{\sqrt{A(r)}}\,\partial _{t}\psi_{\uparrow} - H \sqrt{B(r)}\,\partial_{r}\psi_{\uparrow}\right) + Y m = & 0, \quad 
-\frac{H}{r}\left(\partial _{\theta }\psi_{\uparrow} + \frac{i}{\sin \theta }\,\partial _{\varphi }\psi_{\uparrow}\right) = 0,
\end{split}
\fe%
at the leading order in \(\hbar\), assuming the action is represented as:
$
\psi_{\uparrow}=- \omega\, t + \chi(r) + L(\theta ,\varphi )  $
ensuring that the following equations are obtained:
\cite{vanzo2011tunnelling}
\begin{eqnarray}
\left( \frac{i\, \omega\, H}{\sqrt{A(r)}} - Y \sqrt{B(r)}\, \chi^{\prime }(r)\right) +m\, H &=&0,
\label{bhspin5} \\
-\frac{H}{r}\left( L_{\theta }+\frac{i}{\sin \theta }L_{\varphi }\right) &=&0,
\label{bhspin6} \\
-\left( \frac{i\,\omega\, Y}{\sqrt{A(r)}} + H\sqrt{B(r)}\,  \chi^{\prime }(r)\right) + Y \,m &=&0,
\label{bhspin7} \\
-\frac{H}{r}\left( L_{\theta } + \frac{i}{\sin \theta }L_{\varphi }\right) &=& 0.
\label{bhspin8}
\end{eqnarray}

The forms of \(H\) and \(Y\) do not influence the result that Eqs. (\ref{bhspin6}) and (\ref{bhspin8}) enforce the condition \(L_{\theta} + i(\sin \theta)^{-1} L_{\varphi} = 0\), indicating that \(L(\theta, \varphi)\) must be complex. This requirement is valid for both outgoing and incoming cases. As a result, when the ratio of outgoing to incoming probabilities is determined, the contributions from \(L\) cancel out. This allows \(L\) to be excluded from further calculations. In the massless case, Eqs. (\ref{bhspin5}) and (\ref{bhspin7}) give rise to two distinct solutions:
\ie
H = -i Y, \qquad \chi^{\prime }(r) = \chi_{\text{out}}' = \frac{\omega}{\sqrt{A(r)B(r)}},
\fe
\ie
H = i Y, \qquad \chi^{\prime }(r) = \chi_{\text{in}}' = - \frac{\omega}{\sqrt{A(r)B(r)}},
\fe
where \(\chi_{\text{out}}\) and \(\chi_{\text{in}}\) correspond to the outgoing and incoming solutions, respectively \cite{vanzo2011tunnelling}. As a result, the total tunneling probability is expressed as \(\Gamma_{\psi} \sim e^{-2 \, \text{Im} \, (\chi_{\text{out}} - \chi_{\text{in}})}\). Thus,
\begin{equation}
\mathcal \chi_{ \text{out}}(r)= - \mathcal \chi_{ \text{in}} (r) = \int \mathrm{d} r \,\frac{\omega}{\sqrt{A(r)B(r)}}\:.
\end{equation}

Notably, the dominant energy condition, combined with the Einstein field equations, ensures that \(A(r)\) and \(B(r)\) vanish at the same points. Close to \(r = r_{h}\), these functions can be approximated linearly as:
\ie
A(r)B(r) = A'(r_{h})B'(r_{h})(r - r_{h})^2 + \dots
\fe
indicating the existence of a simple pole with a defined coefficient. Applying Feynman’s method, we find
\ie
2\mbox{Im}\;\left( \mathcal \chi_{ \text{out}} - \mathcal \chi_{ \text{in}} \right) =\mbox{Im}\int \mathrm{d} r \,\frac{4\omega}{\sqrt{A(r)B(r)}}=\frac{2\pi\omega}{\kappa_{+}},
\fe
where the surface gravity is given by \(\kappa_{+} = \frac{1}{2} \sqrt{A'(r_{h}) B'(r_{h})}\). In this framework, the particle density \(n_{\psi}\) for this black hole solution is determined by the expression \(\Gamma_{\psi} \sim e^{-\frac{2 \pi \omega}{\kappa_{+}}}\)
\ie
n_{\psi} = \frac{1}{ e^{\frac{12 \pi  \omega }{\sqrt{\frac{\left(6 M r_{h} + 2 \Lambda  r_{h}^4+3 \xi  r_{h}^3 \mathfrak{Q}-6 \mathfrak{Q}\right)^2}{r_{h}^6}}}}+1}.
\fe

\begin{figure}
    \centering
      \includegraphics[scale=0.61]{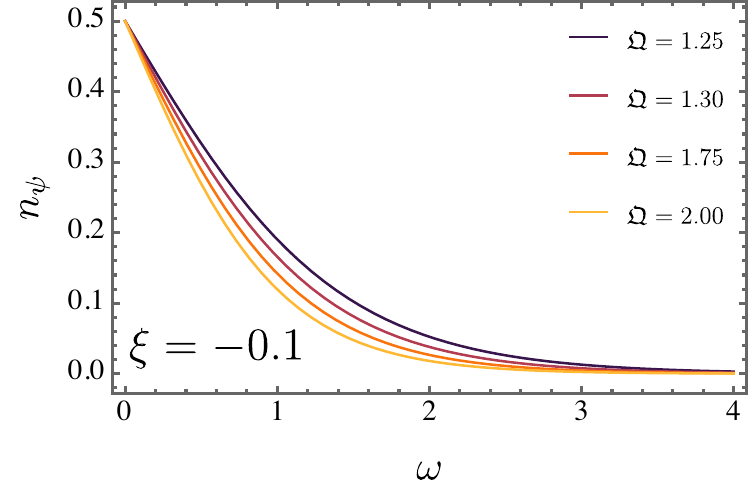}
      \includegraphics[scale=0.61]{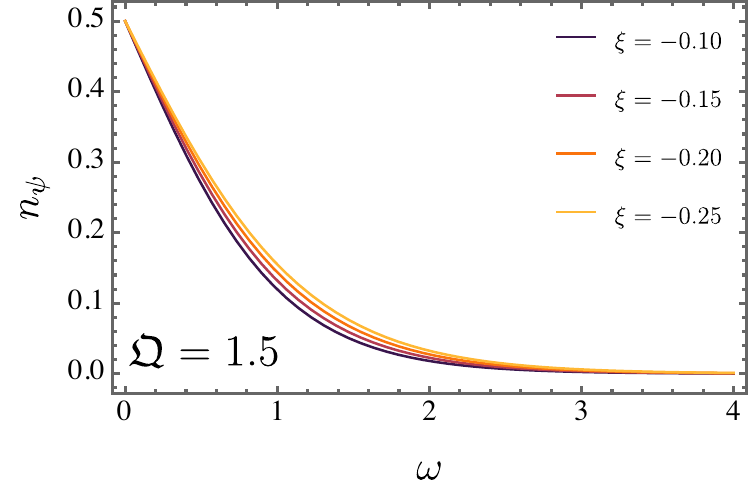}
    \caption{The particle density \( n_{\psi} \) is shown for various values of $\mathfrak{Q}$ (on the left) and $\xi$ (on the right).}
    \label{fermions}
\end{figure}

Fig. \ref{fermions} depicts the behavior of \(n_{\psi}\) as a function of \(\mathfrak{Q}\) and \(\xi\). Generally, increasing \(\mathfrak{Q}\) results in a decrease in the particle density \(n_{\psi}\). Conversely, lowering the values of \(\xi\) leads to an increase in \(n_{\psi}\).

\section{Black hole evaporation}

In this section, we explore the evaporation process of the black hole as it nears its concluding phase. We start by defining the Hawking temperature in terms of the mass \( M \). To achieve this, we substitute the expression for $r_{h}$ given in Eq. (\ref{eventttt}) into Eq. (\ref{hawkingtemmp}), resulting in:
\ie
\begin{split}
T & = \frac{\mathfrak{Q}}{4 \pi  \left(\lambda -\frac{\sqrt{\frac{2\ 2^{2/3} \Lambda  \Xi -24 \Lambda +\frac{36 \sqrt[3]{2} \Lambda  (6 M \xi  \mathfrak{Q}+4 \Lambda  \mathfrak{Q}+1)}{\Xi }+27 \xi ^2 \mathfrak{Q}^2}{\Lambda ^2}}}{4 \sqrt{3}}+\frac{3 \xi  \mathfrak{Q}}{4 \Lambda }\right)^3}+\frac{\xi  \mathfrak{Q}}{2 \pi } \\ 
& -\frac{1}{4 \pi  \left(\lambda -\frac{\sqrt{\frac{2\ 2^{2/3} \Lambda  \Xi -24 \Lambda +\frac{36 \sqrt[3]{2} \Lambda  (6 M \xi  \mathfrak{Q}+4 \Lambda  \mathfrak{Q}+1)}{\Xi }+27 \xi ^2 \mathfrak{Q}^2}{\Lambda ^2}}}{4 \sqrt{3}}+\frac{3 \xi  \mathfrak{Q}}{4 \Lambda }\right)}\\
& +\frac{-12 \lambda  \Lambda +\Lambda  \sqrt{\frac{6 \Lambda  \left(\Xi  \left(2^{2/3} \Xi -12\right)+18 \sqrt[3]{2}\right)+216 \sqrt[3]{2} \Lambda  \mathfrak{Q} (2 \Lambda +3 M \xi )+81 \xi ^2 \Xi  \mathfrak{Q}^2}{\Lambda ^2 \Xi }}-9 \xi  \mathfrak{Q}}{48 \pi },
\end{split}
\fe
where $\Xi$ is given by
\ie
\begin{split}
\nonumber
\Xi  = &\left\{ 972 \Lambda  M^2 \right. \\
& \left. +\sqrt{\left(972 \Lambda  M^2+486 M \xi  \mathfrak{Q}+729 \xi ^2 \mathfrak{Q}^3-648 \Lambda  \mathfrak{Q}+54\right)^2-4 (54 M \xi  \mathfrak{Q}+36 \Lambda  \mathfrak{Q}+9)^3} \right. \\
& \left. +486 M \xi  \mathfrak{Q}+729 \xi ^2 \mathfrak{Q}^3-648 \Lambda  \mathfrak{Q}+54\right\}^{1/3},
\end{split}
\fe
and
\ie
\begin{split}
\lambda  = & \frac{1}{2} \left\{-\frac{\Xi }{3 \sqrt[3]{2} \Lambda }-\frac{4}{\Lambda }+\frac{\sqrt{3} \left(48 \Lambda ^2 M-27 \xi ^3 \mathfrak{Q}^3+36 \Lambda  \xi  \mathfrak{Q}\right)}{2 \Lambda ^3 \sqrt{\frac{2\ 2^{2/3} \Lambda  \Xi -24 \Lambda +\frac{36 \sqrt[3]{2} \Lambda  (6 M \xi  \mathfrak{Q}+4 \Lambda  \mathfrak{Q}+1)}{\Xi }+27 \xi ^2 \mathfrak{Q}^2}{\Lambda ^2}}} \right. \\
& \left. -\frac{3 \sqrt[3]{2} (6 M \xi  \mathfrak{Q}+4 \Lambda  \mathfrak{Q}+1)}{\Lambda  \Xi }+\frac{9 \xi ^2 \mathfrak{Q}^2}{2 \Lambda ^2}\right\}^{1/2}    
\end{split}
\fe

It is important to note that, while we have included the cosmological constant \(\Lambda\), a similar issue arises here, as previously discussed in the literature \cite{AraujoFilho:2024xhm}, when considering the Hawking temperature as a function of mass --- at least for the parameter values used in this study. Interestingly, whether or not the cosmological constant is present, the black hole appears to absorb radiation rather than emit it.

With these properties established, we turn our attention to another key aspect requiring investigation: the lifetime of the black hole. To explore this, we represent
\ie
\frac{\mathrm{d}M}{\mathrm{d}\tau} = - \alpha \sigma a T^{4}.
\fe
Here, \( a \) denotes the radiation constant, \( \sigma \) represents the cross--sectional area, and \( \alpha \) is the greybody factor. It should be highlighted that in future work, we intend to provide an examination of this element along with other characteristics, including gravitational lensing. Additionally, within the geometric optics framework, \( \sigma \) corresponds to the photon capture cross--section, expressed as \(\pi \mathfrak{R}^2\). Consequently, we derive
\ie
\begin{split}
\frac{\mathrm{d}M}{\mathrm{d}\tau} = & - \chi \, \Xi^{2} \left( \frac{-\frac{2 M}{r_{o}}+\frac{\Lambda  r_{o}^2}{3}+\frac{\mathfrak{Q}}{r_{o}^2}+\xi  r_{o} \mathfrak{Q}+1}{\frac{\Lambda  \Xi ^2}{3}-\frac{2 M}{\Xi }+\mathfrak{Q} \left(\xi  \Xi +\frac{1}{\Xi ^2}\right)+1} \right)  \\
& \times  \left\{ \frac{\mathfrak{Q}}{4 \pi  \left(\lambda -\frac{\sqrt{\frac{2\ 2^{2/3} \Lambda  \Xi -24 \Lambda +\frac{36 \sqrt[3]{2} \Lambda  (6 M \xi  \mathfrak{Q}+4 \Lambda  \mathfrak{Q}+1)}{\Xi }+27 \xi ^2 \mathfrak{Q}^2}{\Lambda ^2}}}{4 \sqrt{3}}+\frac{3 \xi  \mathfrak{Q}}{4 \Lambda }\right)^3}+\frac{\xi  \mathfrak{Q}}{2 \pi } \right. \\ 
& \left. -\frac{1}{4 \pi  \left(\lambda -\frac{\sqrt{\frac{2\ 2^{2/3} \Lambda  \Xi -24 \Lambda +\frac{36 \sqrt[3]{2} \Lambda  (6 M \xi  \mathfrak{Q}+4 \Lambda  \mathfrak{Q}+1)}{\Xi }+27 \xi ^2 \mathfrak{Q}^2}{\Lambda ^2}}}{4 \sqrt{3}}+\frac{3 \xi  \mathfrak{Q}}{4 \Lambda }\right)} \right. \\
& \left. +\frac{-12 \lambda  \Lambda +\Lambda  \sqrt{\frac{6 \Lambda  \left(\Xi  \left(2^{2/3} \Xi -12\right)+18 \sqrt[3]{2}\right)+216 \sqrt[3]{2} \Lambda  \mathfrak{Q} (2 \Lambda +3 M \xi )+81 \xi ^2 \Xi  \mathfrak{Q}^2}{\Lambda ^2 \Xi }}-9 \xi  \mathfrak{Q}}{48 \pi } \right\}^{4},
\end{split}
\fe
with $\chi = a \alpha$. In this manner, it yields 
\ie
\begin{split}
\int_{0}^{t_{\text{evap}}} \chi \mathrm{d}\tau & = - \int_{M_{i}}^{M_{f}} 
\left[ \chi \, \Xi^{2} \left( \frac{-\frac{2 M}{r_{o}}+\frac{\Lambda  r_{o}^2}{3}+\frac{\mathfrak{Q}}{r_{o}^2}+\xi  r_{o} \mathfrak{Q}+1}{\frac{\Lambda  \Xi ^2}{3}-\frac{2 M}{\Xi }+\mathfrak{Q} \left(\xi  \Xi +\frac{1}{\Xi ^2}\right)+1} \right) \right.  \\
& \left. \times  \left\{ \frac{\mathfrak{Q}}{4 \pi  \left(\lambda -\frac{\sqrt{\frac{2\ 2^{2/3} \Lambda  \Xi -24 \Lambda +\frac{36 \sqrt[3]{2} \Lambda  (6 M \xi  \mathfrak{Q}+4 \Lambda  \mathfrak{Q}+1)}{\Xi }+27 \xi ^2 \mathfrak{Q}^2}{\Lambda ^2}}}{4 \sqrt{3}}+\frac{3 \xi  \mathfrak{Q}}{4 \Lambda }\right)^3}+\frac{\xi  \mathfrak{Q}}{2 \pi } \right.\right. \\ 
& \left.\left. -\frac{1}{4 \pi  \left(\lambda -\frac{\sqrt{\frac{2\ 2^{2/3} \Lambda  \Xi -24 \Lambda +\frac{36 \sqrt[3]{2} \Lambda  (6 M \xi  \mathfrak{Q}+4 \Lambda  \mathfrak{Q}+1)}{\Xi }+27 \xi ^2 \mathfrak{Q}^2}{\Lambda ^2}}}{4 \sqrt{3}}+\frac{3 \xi  \mathfrak{Q}}{4 \Lambda }\right)} \right.\right. \\
& \left.\left. +\frac{-12 \lambda  \Lambda +\Lambda  \sqrt{\frac{6 \Lambda  \left(\Xi  \left(2^{2/3} \Xi -12\right)+18 \sqrt[3]{2}\right)+216 \sqrt[3]{2} \Lambda  \mathfrak{Q} (2 \Lambda +3 M \xi )+81 \xi ^2 \Xi  \mathfrak{Q}^2}{\Lambda ^2 \Xi }}-9 \xi  \mathfrak{Q}}{48 \pi } \right\}^{4} \right]^{-1} \mathrm{d}M.
\end{split}
\fe
In this scenario, the initial and final black hole masses are designated as \( M_i \) and \( M_f \), respectively, while the evaporation process concludes at \( t_{\text{evap}} \). As suggested by the preceding expression, an analytical solution is not available, necessitating a numerical method. This approach, however, yields an unexpected result: for most parameter values used here, the evaporation time appears as a non--real quantity. Yet, by setting \( M_f = 10 \), \( M_i = 1 \), \( \mathfrak{Q} = 0.5 \), \( \xi = -0.01 \), \( \Lambda = 10^{-5} \), and \( r_o = 100 \), the evaporation time becomes real, specifically \( t_{\text{evap}} = 2.21334 \times 10^{-11} \), indicating a remarkably very short lifetime.


\section{Decaying Quasiresonances}

\subsection{Scalar perturbations}

In the final stages of a black hole’s evolution, a set of oscillations emerges, referred to as quasinormal modes, which encapsulate the black hole’s inherent properties along with the characteristics of the surrounding spacetime. Unlike normal modes that occur in closed systems, quasinormal modes describe open systems where energy dissipates through gravitational waves \cite{Li:2024npg,Daghigh:2024wcl,Daghigh:2023ixh,Daghigh:2022uws,Daghigh:2021psm}. These oscillations are unaffected by the initial disturbances that triggered them and are mathematically represented as poles in the complex Green's function, serving as solutions to the wave equation in the black hole’s background spacetime \cite{Konoplya:2013rxa,Konoplya:2019hlu,Kokkotas:2010zd,Konoplya:2007zx,Konoplya:2011qq}. Calculating these frequencies is often complex, as they depend on solving wave equations in a curved spacetime defined by the metric \( g_{\mu\nu} \) \cite{araujo2023analysis,heidari2023gravitational,araujo2024gravitational}. 

Although exact solutions are generally challenging to obtain, some specific cases, dealing with exact solutions, such as the Schwarzschild--AdS case, have been addressed in the literature (e.g., Ref. \cite{senjaya2024scalar}), approximation techniques like the Wentzel--Kramers--Brillouin (WKB) method, initially proposed by Will and Iyer \cite{iyer1987black,iyer1987black1} and refined by Konoplya \cite{konoplya2003quasinormal}, provide effective tools. This method is applied in our work to study scalar field perturbations by solving the Klein--Gordon equation, namely, $
\frac{1}{\sqrt{-g}}\partial_{\mu}(g^{\mu\nu}\sqrt{-g}\partial_{\nu}\Phi) = 0$, within curved spacetime to extract the quasinormal frequencies. Here, we concentrate on analyzing the scalar field as a small perturbation. In this manner, the Klein--Gordon equation reads 
\ie
\label{jekjsk}
\begin{split}
-& \frac{1}{f(r)} \frac{\partial^{2} \Phi}{\partial t^{2}} + \frac{1}{r^{2}} \left[  \frac{\partial}{\partial r} \left(  f(r) \, r^{2}  \frac{\partial \Phi}{\partial r}  \right)  \right] \\  + & \frac{1}{r^{2} \sin \theta}  \left[  \frac{\partial }{\partial \theta} \left( \sin \theta \frac{\partial}{\partial \theta} \Phi   \right)        \right] 
 +  \frac{1}{r^{2} \sin^{2}}  \frac{\partial^{2} \Phi}{\partial \phi^{2}} = 0,
\end{split}
\fe
where $\sqrt{-g} = r^{2}\sin\theta$. Due to the spherical symmetry of the system, the scalar field can be decomposed as 
\ie
\Phi(t, r, \theta, \varphi) = \sum_{l=0}^{\infty} \sum_{m=-l}^{l}  Y_{lm}(\theta, \varphi) \frac{\Psi(t,r)}{r} ,
\fe
where \(Y_{lm}(\theta, \varphi)\) represents the spherical harmonics. With this decomposition, the radial part of Eq. (\ref{jekjsk}) can be rewritten as 
\ie
\frac{\partial^{2}\Psi(r)}{\partial t^{2}}  + \frac{f(r)}{r} \left\{ \frac{\partial }{\partial r}  \left[ f(r) r^{2} \frac{\partial}{\partial r}  \left( \frac{\Psi(r)}{r} \right)   \right]     \right\} - f(r) \frac{\ell(\ell + 1)}{r^{2}}\Psi(r) = 0 .
\fe

Now, let us introduce the tortoise coordinate \(r^*\), defined by \(\mathrm{d}r^* = \frac{1}{\sqrt{f(r)^2}} \, \mathrm{d}r\), which reads $\int \mathrm{d}r^* =  \int 1/[ 1 - \frac{2M}{r} + \frac{\mathfrak{Q}}{r^{2}}(1 + \xi r^{3})  + \frac{\Lambda}{3}r^{2}] \, \mathrm{d}r $ and, therefore, 
\ie
\begin{split}
r^{*} = \, \, & r + \frac{r_{1}^4 \ln (r-r_{1})}{(r_{1}-r_{2}) (r_{1} - r_{3}) (r_{1} - r_{h})}-\frac{r_{2}^4 \ln (r - r_{2})}{(r_{1} - r_{2}) (r_{2} - r_{3}) (r_{2} - r_{h})} \\
& + \frac{r_{3}^4 \ln (r - r_{3})}{(r_{1} - r_{3}) (r_{2} - r_{3}) (r_{3} - r_{h})}-\frac{r_{h}^4 \ln (r - r_{h})}{(r_{1} - r_{h}) (r_{2}-r_{h}) (r_{3} - r_{h})}.
\end{split}
\fe

Thereby, Klein--Gordon equation transforms it into a Schrödinger--like equation
\ie
-\frac{\partial^2 \Psi(r)}{\partial t^2} + \frac{\partial^2 \Psi(r)}{\partial r^{*2}} + \mathrm{V}(r) \Psi(r) = 0.
\fe

\begin{figure}
    \centering
    \includegraphics[scale=0.6]{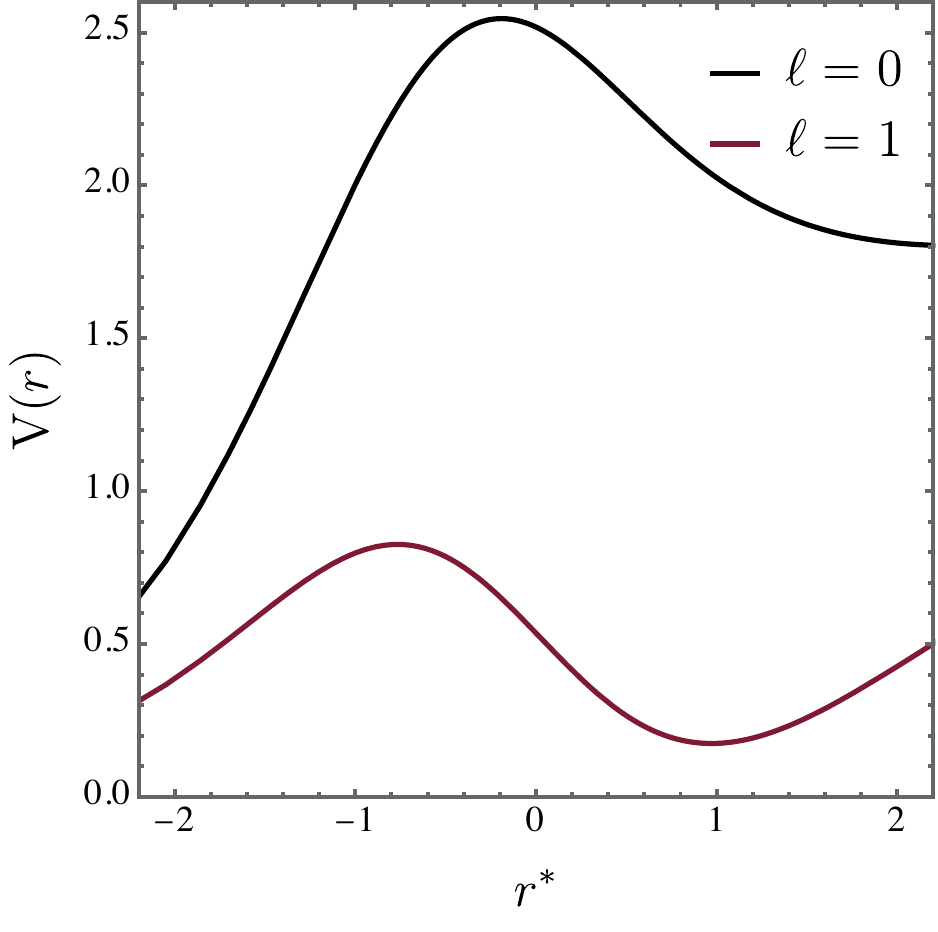}
    \caption{The effective potential \(\mathrm{V}(r)\) is shown as a function of the tortoise coordinate \(r^{*}\) for scalar perturbations, with varying values of \(\ell\).}
    \label{sdffsa}
\end{figure}

Here, \(\mathrm{V}(r)\), known as the Regge--Wheeler potential, encodes key details about the black hole’s geometry. 
It should be noted that, depending on the particular values of the parameters \(\mathfrak{Q}\), \(\Lambda\), \(M\), and \(\xi\), the horizons $r_{1}, r_{2}$ and $r_{3}$ may no longer hold physical relevance. In addition, the effective potential can be written as follows
\ie
\begin{split}
\mathrm{V}(r) = f(r) \left[ \frac{\ell (\ell+1)}{r^2}+\frac{\frac{2 M}{r^2}-\frac{2 \mathfrak{Q} \left(\xi  r^3+1\right)}{r^3}+\frac{2 \Lambda  r}{3}+3 \xi  \mathfrak{Q}}{r} \right]
\end{split},
\fe
in which $f(r) \equiv \left[  1 - \frac{2M}{r} + \frac{\mathfrak{Q}}{r^{2}}(1 + \xi r^{3}) +\frac{\Lambda }{3} r^2   \right]$. 
Fig. \ref{sdffsa} presents the effective potential \(\mathrm{V}(r)\) plotted against the tortoise coordinate \(r^{*}\), considering various values of \(\ell\). In this scenario, the bell--shaped curve is clearly evident, indicating that the WKB method can be appropriately applied. This characteristic will be examined in detail in the following subsection.

\begin{table}[!h]
\begin{center}
\caption{\label{qnmtac0} The table displays the quasinormal modes for \( \ell = 0 \) as a function of the parameters \(\xi\) and \(\mathfrak{Q}\) (when $\Lambda = 10^{-5}$) for scalar perturbations.}
\begin{tabular}{c| c | c | c} 
 \hline\hline\hline 
 \,\, $\xi$ \,\,\,\,\,\,  $\mathfrak{Q}$  & $\omega_{0}$ & $\omega_{1}$ & $\omega_{2}$  \\ [0.2ex] 
 \hline 
 -0.01,  1.0  & 0.303245 - 0.0347466$i$  & 0.988177 - 0.0254463$i$ & 2.58650 - 0.0222081$i$   \\
 
 -0.02,  1.0  & 0.223920 - 0.0424307$i$ & 0.680039 - 0.0340783$i$  &  1.82470 - 0.0292125$i$   \\
 
 -0.03,  1.0  & 0.490987 - 0.0172516$i$  & 1.726020 - 0.0120182$i$ &  4.43977 - 0.0103675$i$   \\
 
 -0.04, 1.0  & 0.501856 - 0.0145696$i$ & 1.782930 - 0.0091120$i$ & 4.59675 - 0.0056717$i$ \\
 
 -0.05, 1.0  &  0.013014 - 0.5178120$i$ & 0.0092842 - 1.884150$i$ & 0.00760 - 4.7845700$i$  \\
   [0.2ex] 
 \hline\hline\hline 
 \,\, $\xi$ \,\,\,\,\,\,  $\mathfrak{Q}$  & $\omega_{0}$ & $\omega_{1}$ & $\omega_{2}$  \\ [0.2ex] 
 \hline 
 -0.01,  0.6  & 0.121137 - 0.100181$i$   & 0.104823 - 0.325132$i$  & 0.172124 - 0.469685$i$  \\
 
 -0.01,  0.7  & 0.119105 - 0.102930$i$ & 0.095724 - 0.353806$i$  &  0.119154 - 0.618653$i$  \\

 -0.01,  0.8  & 0.132726 - 0.090526$i$  & 0.114700 - 0.268089$i$ &  0.171936 - 0.319567$i$ \\
 
 -0.01, 0.9  & 0.118550 - 0.093550$i$ & 0.065365 - 0.379798$i$ & 0.045669 - 0.826877$i$ \\
 
 -0.01, 1.0  & 0.303245 - 0.034746$i$  & 0.988177 - 0.025446$i$ & 2.586500 - 0.022208$i$ \\
   [0.2ex] 
 \hline \hline \hline 
\end{tabular}
\end{center}
\end{table}

\begin{table}[!h]
\begin{center}
\caption{\label{qnmtac1} The table displays the quasinormal modes for \( \ell = 1 \) as a function of the parameters \(\xi\) and \(\mathfrak{Q}\) (when $\Lambda = 10^{-5}$) for scalar perturbations.}
\begin{tabular}{c| c | c | c} 
 \hline\hline\hline 
 \,\, $\xi$ \,\,\,\,\,\,  $\mathfrak{Q}$  & $\omega_{0}$ & $\omega_{1}$ & $\omega_{2}$  \\ [0.2ex] 
 \hline 
 -0.01,  1.0  & 0.358881 - 0.0863109$i$  & 0.312800 - 0.282487$i$ & 0.227504 - 0.585939$i$  \\
 
 -0.02,  1.0  & 0.345688 - 0.0817201$i$ & 0.330598 - 0.244330$i$  &  0.314978 - 0.388265$i$  \\
 
 -0.03,  1.0  & 0.327740 - 0.0780300$i$  & 0.308431 - 0.237781$i$ &  0.274830 - 0.406268$i$  \\
 
 -0.04, 1.0  & 0.308933 - 0.0742770$i$  & 0.281787 - 0.234226$i$ & 0.226327 - 0.446558$i$ \\
 
 -0.05, 1.0  &  0.290080 - 0.0702429$i$ & 0.258939 - 0.227078$i$  & 0.196797 - 0.460718$i$  \\
   [0.2ex] 
 \hline\hline\hline 
 \,\, $\xi$ \,\,\,\,\,\,  $\mathfrak{Q}$  & $\omega_{0}$ & $\omega_{1}$ & $\omega_{2}$  \\ [0.2ex] 
 \hline 
 -0.01,  0.6  & 0.322293 - 0.0962351$i$ &  0.299903 - 0.298481$i$ & 0.272390 - 0.520571$i$   \\
 
 -0.01,  0.7  & 0.329817 - 0.0952800$i$ & 0.308720 - 0.294588$i$  & 0.281913 - 0.511660$i$  \\

 -0.01,  0.8  & 0.338625 - 0.0936982$i$  & 0.318407 - 0.288632$i$ &  0.290542 - 0.499476$i$   \\
 
 -0.01, 0.9  & 0.349148 - 0.0909009$i$ & 0.328800 - 0.278382$i$ & 0.296372 - 0.478725$i$  \\
 
 -0.01,  1.0  & 0.358881 - 0.0863109$i$  & 0.312800 - 0.282487$i$ & 0.227504 - 0.585939$i$  \\
   [0.2ex] 
 \hline \hline \hline 
\end{tabular}
\end{center}
\end{table}

At this point, the goal is to obtain stationary solutions by assuming the wave function \(\Psi(t,r)\) takes the form \(\Psi(t,r) = e^{-i\omega t} \varphi(r)\), where \(\omega\) denotes the frequency. This assumption simplifies the problem, separating the time--dependent part and reducing to \(\frac{\partial^{2} \varphi}{\partial r^{*2}} - \left[\omega^{2} - \mathrm{V}(r^{*})\right]\varphi = 0\). The appropriate boundary conditions must be applied, requiring that the solution consists of purely ingoing waves at the event horizon to maintain physical consistency. In other words, we have
\[
    \varphi(r^{*}) \sim 
\begin{cases}
    \tilde{\vartheta}_{\ell}(\omega) e^{-i\omega r^{*}} & ( r^{*}\rightarrow - \infty)\\
    \vartheta^{(-)}_{\ell}(\omega) e^{-i\omega r^{*}} + \vartheta^{(+)}_{\ell}(\omega) e^{+i\omega r^{*}} & (r^{*}\rightarrow + \infty).
\end{cases}
\] 
The constants \(\tilde{\vartheta}_\ell(\omega)\), \(\vartheta^{(-)}_\ell(\omega)\), and \(\vartheta^{(+)}_\ell(\omega)\) are crucial for identifying the quasinormal modes. These modes are found by imposing the condition \(\vartheta^{(-)}_\ell(\omega_{n\ell}) = 0\), resulting in outgoing waves at infinity and ingoing waves at the horizon. Here, \(n\) and \(\ell\) represent the overtone and multipole numbers.

The QNM frequencies are derived from the eigenvalues of the time--independent equation, and the WKB approximation is used to compute these frequencies. This semi--analytical technique, originally introduced by Schutz and Will \cite{schutz1985black}, and later improved by Konoplya \cite{konoplya2003quasinormal, konoplya2004quasinormal}, is particularly effective when the potential takes on a barrier--like shape. By expanding the solution near the potential’s maximum, the quasinormal frequencies can be calculated with precision. The final formula for the frequencies is 
$\frac{i(\omega^{2}_{n}-V_{0})}{\sqrt{-2 V^{''}_{0}}} - \sum^{6}_{j=2} \Lambda_{j} = n + \frac{1}{2},$
where \(V''_0\) is the second derivative of the potential at its maximum, and the constants \(\Lambda_j\) (where it represents the higher--order correction terms in the WKB approximation) depend on the potential and its derivatives at this point, playing a crucial role in the accurate determination of the quasinormal mode frequencies.

Tabs. \ref{qnmtac0} and \ref{qnmtac1} display the quasinormal frequencies for various values of \(\xi\) and \(\mathfrak{Q}\) (with \(\Lambda = 10^{-5}\)), emphasizing the cases \(\ell=0\) and \(\ell=1\), respectively. This approach enables the identification of key characteristics of the generalized non--linear electromagnetic solution. Notably, an increase in the effective charge \(\mathfrak{Q}\) results in reduced damping of scalar waves. Additionally, Fig. \ref{convergences} shows the convergence of higher--order corrections in the WKB method, highlighting the precision of the technique used in this analysis.

\begin{figure}
    \centering
    \includegraphics[scale=0.5]{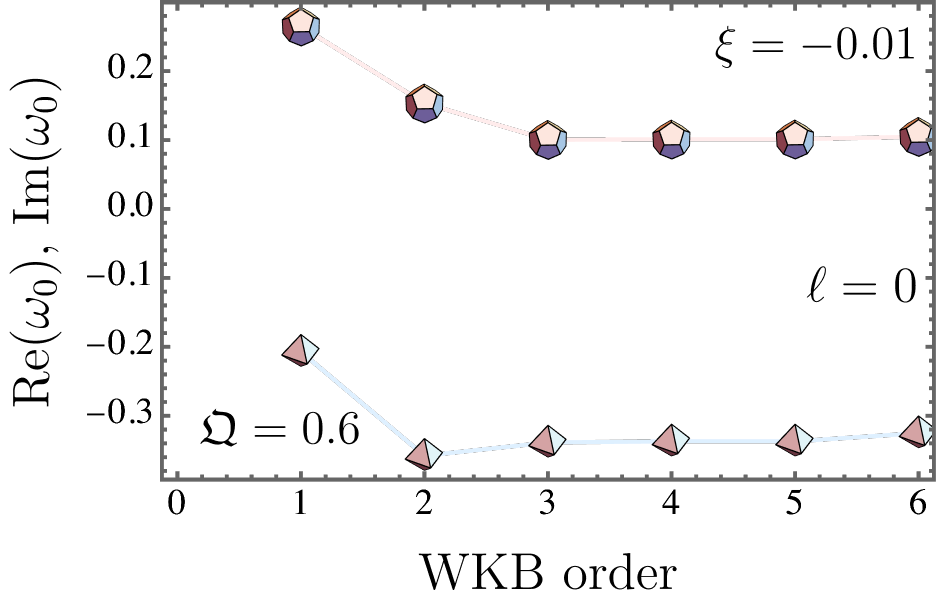}
    \includegraphics[scale=0.51]{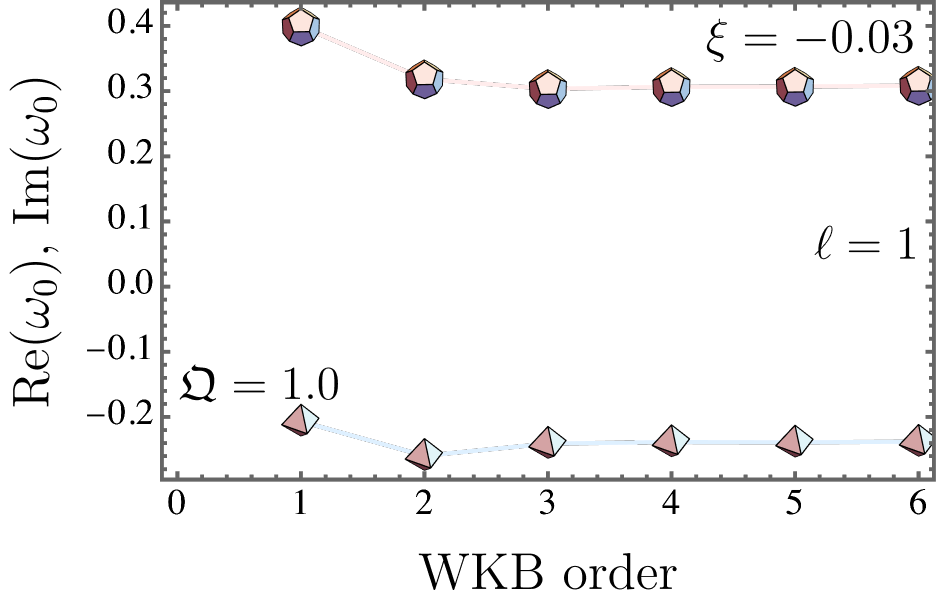}
    \caption{The higher--order convergence of the WKB method for scalar perturbations.}
    \label{convergences}
\end{figure}


\subsection{Vectorial perturbations}

Next, we consider the electromagnetic perturbation, requiring the use of the conventional tetrad formalism \cite{chandrasekhar1998mathematical, Bouhmadi-Lopez:2020oia, Gogoi:2023kjt}, where a basis \(\mathfrak{e}_\mu^{a}\) is established in relation to the black hole metric \(g_{\mu\nu}\). This chosen basis adheres to the conditions
\ie
\mathfrak{e}^{a}_\mu \mathfrak{e}^\mu_{b} = \delta^{a}_{b}, \, \, \, \,
\mathfrak{e}^{a}_\mu \mathfrak{e}^\nu_{a} = \delta^{\nu}_{\mu}, \, \, \, \,
\mathfrak{e}^{a}_\mu = g_{\mu\nu} \mathfrak{n}^{a b} \mathfrak{e}^\nu_{b}, \, \, \, \,
g_{\mu\nu} = \mathfrak{n}_{a b}\mathfrak{e}^{a}_\mu \mathfrak{e}^{b}_\nu = \mathfrak{e}_{a\mu} \mathfrak{e}^{a}_\nu.
\fe

In the context of electromagnetic perturbations within the tetrad formalism, the Bianchi identity for the field strength, \(\mathfrak{F}_{[ab|c]} = 0\), leads to
\begin{align}
\left( r \sqrt{g_{tt}}\, \mathfrak{F}_{t \phi}\right)_{,r} + r \sqrt{g_{rr}}\,
\mathfrak{F}_{\phi r, t} &=0,  \label{edem1} \\
\left( r \sqrt{g_{tt}}\, \mathfrak{F}_{ t \phi}\sin\theta\right)_{,\theta} + r^2
\sin\theta\, \mathfrak{F}_{\phi r, t} &=0.  \label{edem2}
\end{align}

Thus, the conservation equation can be expressed below
\ie
\mathfrak{n}^{b c}\! \left( \mathfrak{F}_{a b} \right)_{|c} =0.
\fe

This equation can additionally be reformulated using spherical polar coordinates in the following form:
\ie  \label{edem3}
\left( r \sqrt{g_{tt}}\, \mathfrak{F}_{\phi r}\right)_{,r} + \sqrt{g_{tt} g_{rr}}%
\, \mathfrak{F}_{\phi \theta,\theta} + r \sqrt{g_{rr}}\, \mathfrak{F}_{t \phi, t} = 0.
\fe

In these expressions, the vertical bar and comma indicate intrinsic and directional derivatives with respect to the tetrad indices. By utilizing Eqs. \eqref{edem1} and \eqref{edem2}, together with the time derivative of Eq. 
\eqref{edem3} we obtain 
\ie  \label{edem4}
\left[ \sqrt{g_{tt} g_{rr}^{-1}} \left( r \sqrt{g_{tt}}\, \mathcal{F}
\right)_{,r} \right]_{,r} + \dfrac{g_{tt} \sqrt{g_{rr}}}{r} \left( \dfrac{%
\mathcal{F}_{,\theta}}{\sin\theta} \right)_{,\theta}\!\! \sin\theta - r 
\sqrt{g_{rr}}\, \mathcal{F}_{,tt} = 0,
\fe
where \(\mathcal{F} = \mathfrak{F}_{t \phi } \sin\theta\). Applying Fourier decomposition \((\partial_t \rightarrow - i \omega)\) and decomposing the field as \(\mathcal{F}(r,\theta) = \mathcal{F}(r) Y_{,\theta} / \sin\theta\), where \(Y(\theta)\) represents the Gegenbauer function \cite{g1,g2,g3,g5,g6}, Eq. \eqref{edem4} can be rewritten as follows \footnote{For a comprehensive derivation of this relation, please consult Refs. \cite{g1, g2, g3, g5, g6} and the references cited therein.}:
\ie  \label{edem5}
\left[ \sqrt{g_{tt} g_{rr}^{-1}} \left( r \sqrt{g_{tt}}\, \mathcal{F}
\right)_{,r} \right]_{,r} + \omega^2 r \sqrt{g_{rr}}\, \mathcal{F} -
g_{tt} \sqrt{g_{rr}} r^{-1} \ell(\ell+1)\, \mathcal{F} = 0.
\fe

By defining \(\psi_e \equiv r \sqrt{g_{tt}} \, \mathcal{F}\), Eq. \eqref{edem5} can be rewritten in a Schrödinger--like form. In this manner, it reads
\ie
\partial^2_{r_*} \psi_{\text{e}} + \omega^2 \psi_{\text{e}} = \mathrm{V}_{\text{e}}(r) \psi_{\text{e}},
\fe
so that the effective potential for the vectorial perturbation is
\ie  
\mathrm{V}_{\text{e}}(r) = g_{tt} \, \dfrac{\ell(\ell+1)}{r^2}.
\fe

\begin{figure}
    \centering
    \includegraphics[scale=0.75]{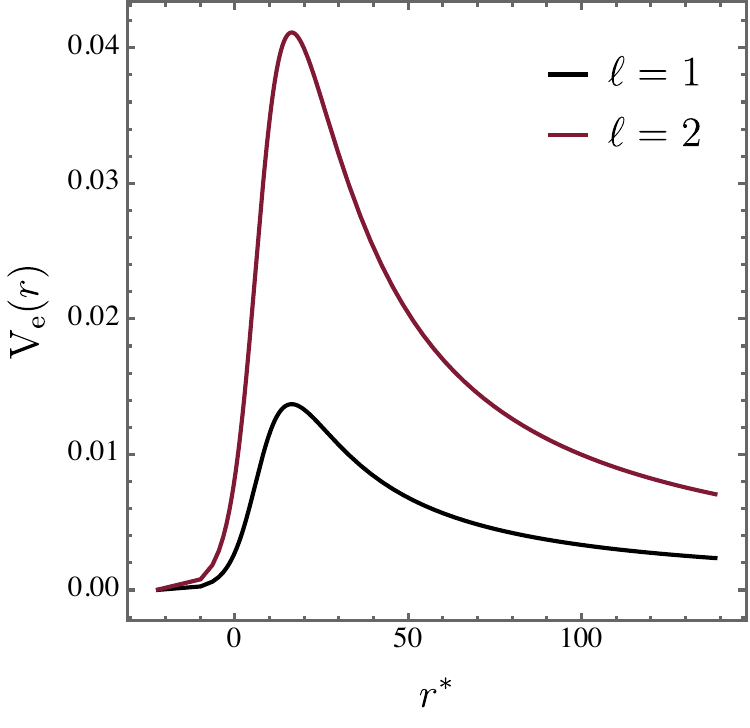}
    \caption{The effective potential $\mathrm{V}_{\text{e}}(r)$ is depicted as a function of the tortoise coordinate $r^{*}$ for vector perturbations, specifically considering different values of $\ell$'s.}
    \label{vectorialv}
\end{figure}

Similarly to our approach for scalar perturbations, we show the vector effective potential $\mathrm{V}_{\text{e}}(r)$ as a function of the tortoise coordinate $r^{*}$ in Fig. \ref{vectorialv} for different configurations of $\ell$. Furthermore, Tabs. \ref{lzerovac} and \ref{lonevac} present the outcomes for vector perturbations. Overall, it is observed that a reduction in \(\xi\) (with \(\mathfrak{Q} = 1.0\) fixed) and an increase in \(\mathfrak{Q}\) (with \(\xi = -0.01\) held constant) lead to oscillations with lower damping.

\begin{figure}
    \centering
    \includegraphics[scale=0.5]{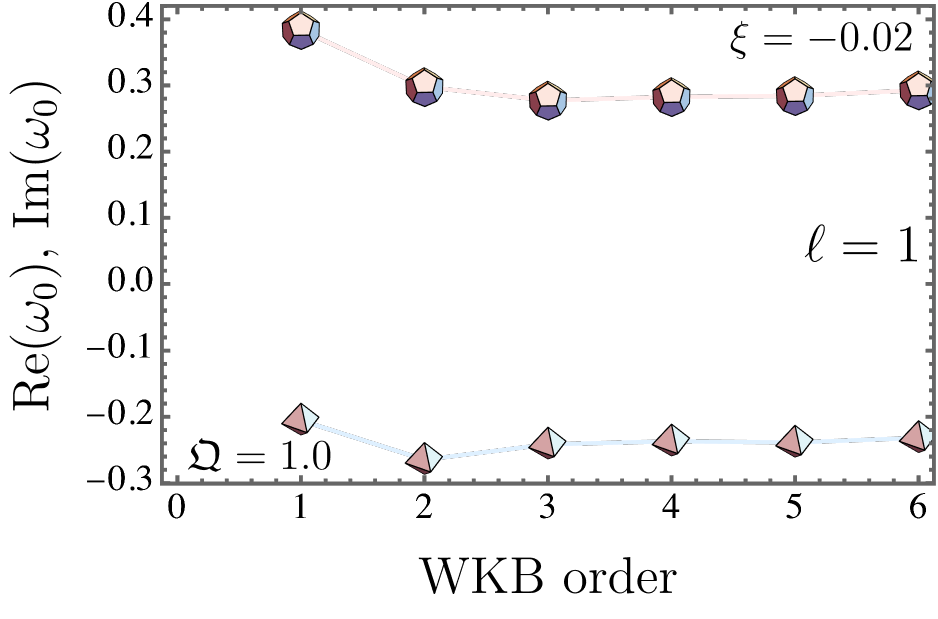}
    \includegraphics[scale=0.51]{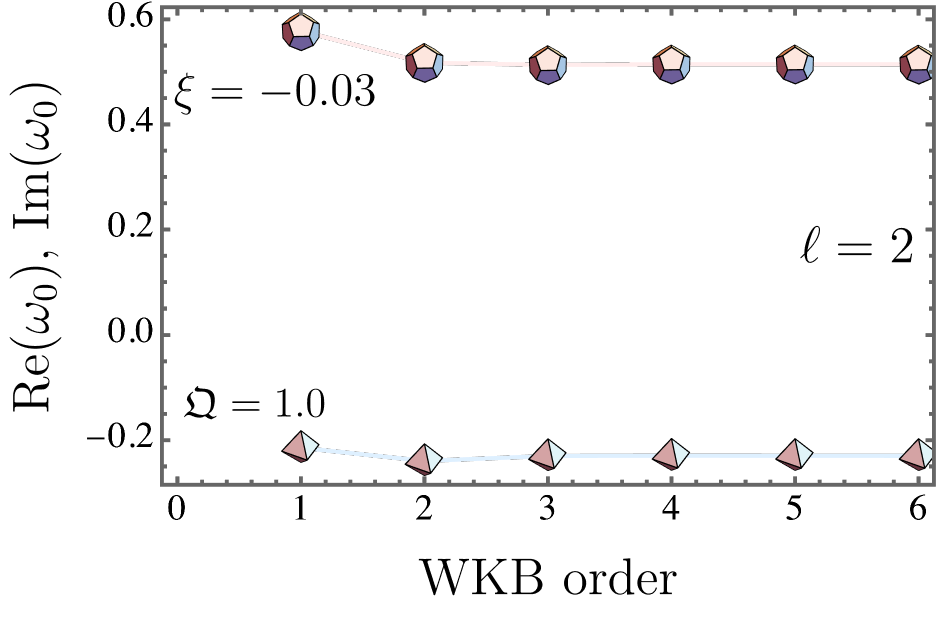}
    \caption{The higher--order convergence of the WKB method for vectorial perturbations.}
    \label{convergences}
\end{figure}

\begin{table}[!h]
\begin{center}
\caption{\label{lzerovac} The table displays the quasinormal modes for \( \ell = 1 \) as a function of the parameters \(\xi\) and \(\mathfrak{Q}\) (when $\Lambda = 10^{-5}$) for vectorial perturbations.}
\begin{tabular}{c| c | c | c} 
 \hline\hline\hline 
 \,\, $\xi$ \,\,\,\,\,\,  $\mathfrak{Q}$  & $\omega_{0}$ & $\omega_{1}$ & $\omega_{2}$  \\ [0.2ex] 
 \hline 
 -0.01,  1.0  & 0.321269 - 0.0815576$i$  & 0.284841 - 0.258698$i$ & 0.217103 - 0.493677$i$   \\
 
 -0.02,  1.0  & 0.309966 - 0.0775775$i$ & 0.292867 - 0.231781$i$  &  0.271185 - 0.366729$i$   \\
 
 -0.03,  1.0  & 0.295462 - 0.0741269$i$  & 0.275842 - 0.225123$i$ &  0.242900 - 0.378265$i$   \\
 
 -0.04, 1.0  & 0.283164 - 0.0698260$i$ & 0.282123 - 0.199615$i$  & 0.317243 - 0.265572$i$ \\
 
 -0.05, 1.0  & 0.264253 - 0.0668523$i$ & 0.233119 - 0.216830$i$ & 0.174718 - 0.437925$i$  \\
   [0.2ex] 
 \hline\hline\hline 
 \,\, $\xi$ \,\,\,\,\,\,  $\mathfrak{Q}$  & $\omega_{0}$ & $\omega_{1}$ & $\omega_{2}$  \\ [0.2ex] 
 \hline 
 -0.01,  0.6  &  0.279277 - 0.0921422$i$   & 0.252778 - 0.288672$i$  & 0.221163 - 0.509954$i$   \\
 
 -0.01,  0.7  & 0.287304 - 0.0913908$i$ & 0.262580 - 0.285210$i$  & 0.232467 - 0.501116$i$  \\

 -0.01,  0.8  & 0.296811 - 0.0899842$i$  & 0.273625 - 0.279537$i$ &  0.243464 - 0.488694$i$ \\
 
 -0.01, 0.9  & 0.308246 - 0.0872432$i$  & 0.284694 - 0.269885$i$  & 0.247968 - 0.472643$i$ \\
 
 -0.01, 1.0  & 0.321269 - 0.0815576$i$  &  0.284841 - 0.258698$i$ & 0.217103 - 0.493677$i$  \\
   [0.2ex] 
 \hline \hline \hline 
\end{tabular}
\end{center}
\end{table}

\begin{table}[!h]
\begin{center}
\caption{\label{lonevac} The table displays the quasinormal modes for \( \ell = 2 \) as a function of the parameters \(\xi\) and \(\mathfrak{Q}\) (when $\Lambda = 10^{-5}$) for vectorial perturbations.}
\begin{tabular}{c| c | c | c} 
 \hline\hline\hline 
 \,\, $\xi$ \,\,\,\,\,\,  $\mathfrak{Q}$  & $\omega_{0}$ & $\omega_{1}$ & $\omega_{2}$  \\ [0.2ex] 
 \hline 
 -0.01,  1.0  & 0.577490 - 0.0834849$i$  & 0.560643 - 0.252862$i$ & 0.527944 - 0.429697$i$   \\
 
 -0.02,  1.0  & 0.552832 - 0.0798297$i$ & 0.536853 - 0.241996$i$   &  0.503295 - 0.413982$i$   \\
 
 -0.03,  1.0  & 0.527682 - 0.0759757$i$  & 0.514155 - 0.229828$i$ &  0.487349 - 0.389866$i$    \\
 
 -0.04, 1.0  & 0.501756 - 0.0719473$i$ & 0.490219 - 0.217335$i$  & 0.468675 - 0.366529$i$ \\
 
 -0.05, 1.0  & 0.475105 - 0.0677144$i$ & 0.466736 - 0.203681$i$  & 0.456129 - 0.336940$i$  \\
   [0.2ex] 
 \hline\hline\hline 
 \,\, $\xi$ \,\,\,\,\,\,  $\mathfrak{Q}$  & $\omega_{0}$ & $\omega_{1}$ & $\omega_{2}$  \\ [0.2ex] 
 \hline 
 -0.01,  0.6  &  0.507642 - 0.0938722$i$   & 0.491105 - 0.285900$i$  & 0.462942 - 0.489290$i$  \\
 
 -0.01,  0.7  & 0.520460 - 0.0930310$i$ & 0.504972 - 0.282962$i$  & 0.478401 - 0.483147$i$  \\

 -0.01,  0.8  & 0.535643 - 0.0915847$i$  & 0.521176 - 0.278077$i$ &  0.495918 - 0.473399$i$ \\
 
 -0.01, 0.9  & 0.554133 - 0.0889419$i$ & 0.540088 - 0.269444$i$ & 0.514304 - 0.457192$i$ \\
 
 -0.01, 1.0  & 0.577490 - 0.0834849$i$  &  0.560643 - 0.252862$i$ & 0.527944 - 0.429697$i$ \\
   [0.2ex] 
 \hline \hline \hline 
\end{tabular}
\end{center}
\end{table}


\section{Time--domain solution}

A comprehensive study of scalar perturbations in the time domain is essential to examine the influence of the quasinormal spectrum on time--dependent scattering phenomena. The complexity of the effective potential, however, demands a precise approach for a more in--depth understanding. For this purpose, we adopt the characteristic integration method introduced by Gundlach et al. \cite{Gundlach:1993tp}, a powerful tool for analyzing this problem. This method allows us to examine the role of quasinormal modes in time--dependent scattering contexts, offering critical perspectives on black holes and related phenomena.

The approach outlined in Refs. \cite{Gundlach:1993tp, Shao:2023qlt, Bolokhov:2024ixe, Guo:2023nkd, Baruah:2023rhd, Skvortsova:2024wly, Yang:2024rms} centers on employing light--cone coordinates, defined as \( u = t - r^{*} \) and \( v = t + r^{*} \). This coordinate framework facilitates a reformulation of the wave equation, making it more conducive to be analyzed. In this way, we write 
\ie
\left(  4 \frac{\partial^{2}}{\partial u \partial v} + V(u,v)\right) \bar{\psi} (u,v) = 0 \label{timedomain}.
\fe

Efficient integration of the expression can be achieved by applying a discretization scheme that incorporates a basic finite--difference method along with numerical techniques
\ie
\bar{\psi}(N) = -\bar{\psi}(S) + \bar{\psi}(W) + \bar{\psi}(E) - \frac{h^{2}}{8}V(S)[\bar{\psi}(W) + \bar{\psi}(E)] + \mathcal{O}(h^{4}),
\fe
in which \( S = (u, v) \), \( W = (u + h, v) \), \( E = (u, v + h) \), and \( N = (u + h, v + h) \), with \( h \) as the grid scaling factor. The null surfaces defined by \( u = u_{0} \) and \( v = v_{0} \) are particularly important, as they establish the locations for setting the initial conditions. In this study, a Gaussian profile centered at \( v = v_{c} \) with a width parameter \( \sigma \) is applied on the null surface \( u = u_{0} \) for our initial data
\ie
\bar{\psi}(u=u_{0},v) = A e^{-(v-v_{0})^{2}}/2\sigma^{2}, \,\,\,\,\,\, \bar{\psi}(u,v_{0}) = \bar{\psi}_{0}.
\fe

Notice that at \( v = v_{0} \), a constant initial condition, \(\bar{\psi}(u, v_{0}) = \bar{\psi}_{0}\), is applied, where, without loss of generality, we set \(\bar{\psi}_{0} = 0\). The integration then advances along lines of constant \( u \) as \( v \) increases, following the specification of the null data. This study presents the results from examining the scalar test field. For simplicity, we assign \( M = 1 \), with the null data characterized by a Gaussian profile centered at \( v = -40 \), having a width of \( \sigma = 1 \), and \(\bar{\psi}_{0} = 0\). The grid covers the intervals \( u \in [120, 200] \) and \( v \in [120, 200] \).

\begin{figure}
    \centering
    \includegraphics[scale=0.45]{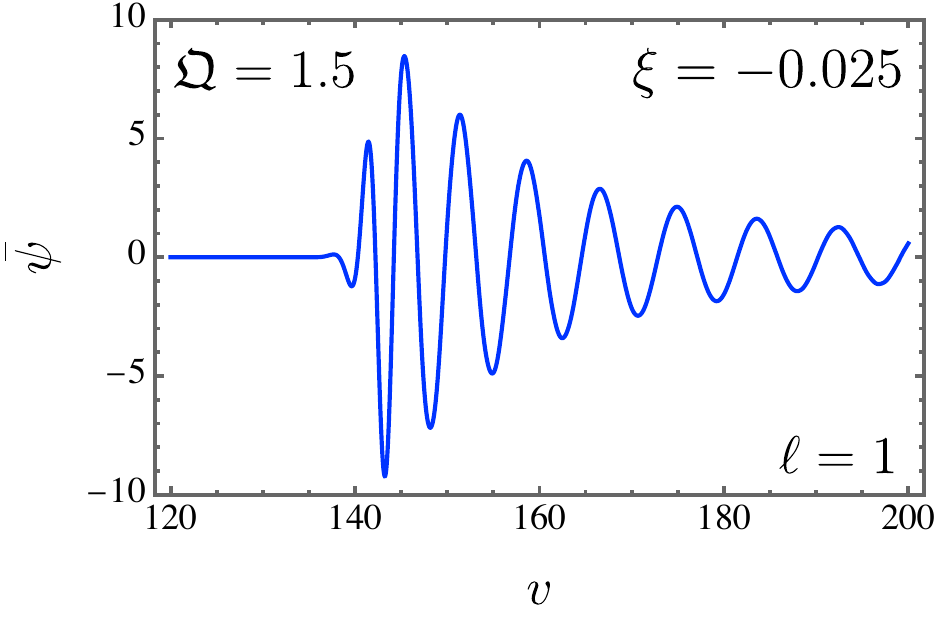}
    \includegraphics[scale=0.45]{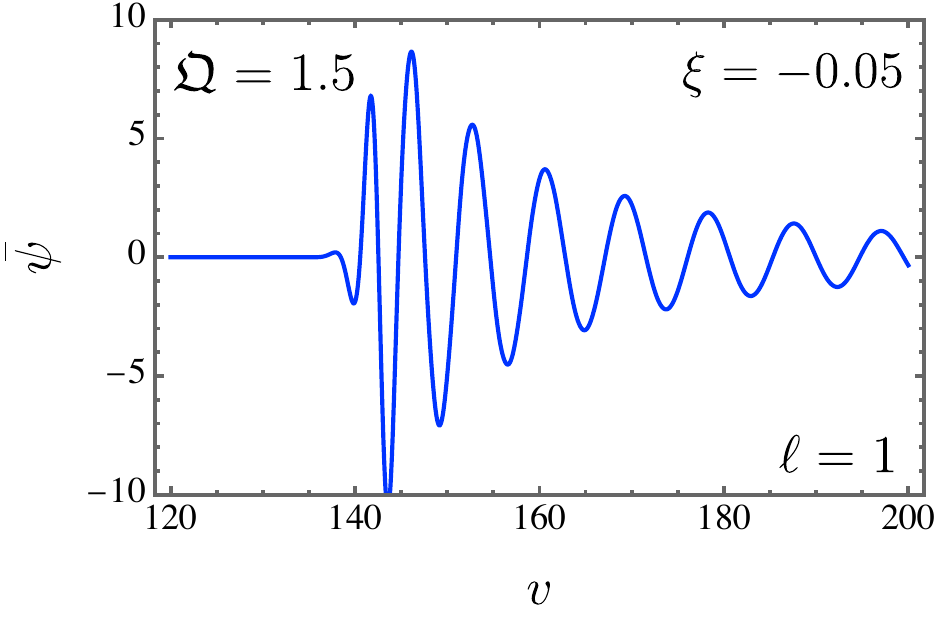}
     \includegraphics[scale=0.45]{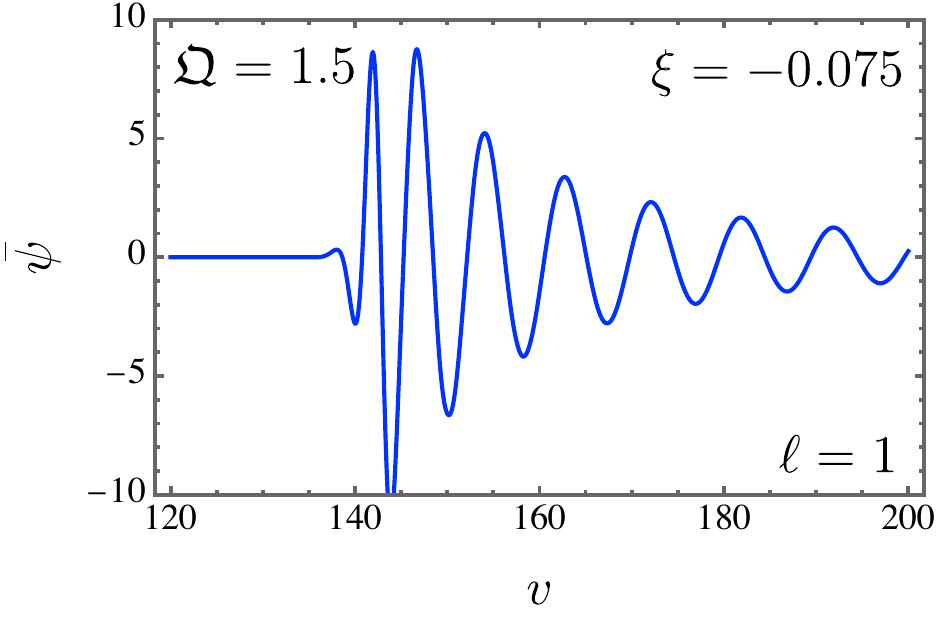}
     \includegraphics[scale=0.45]{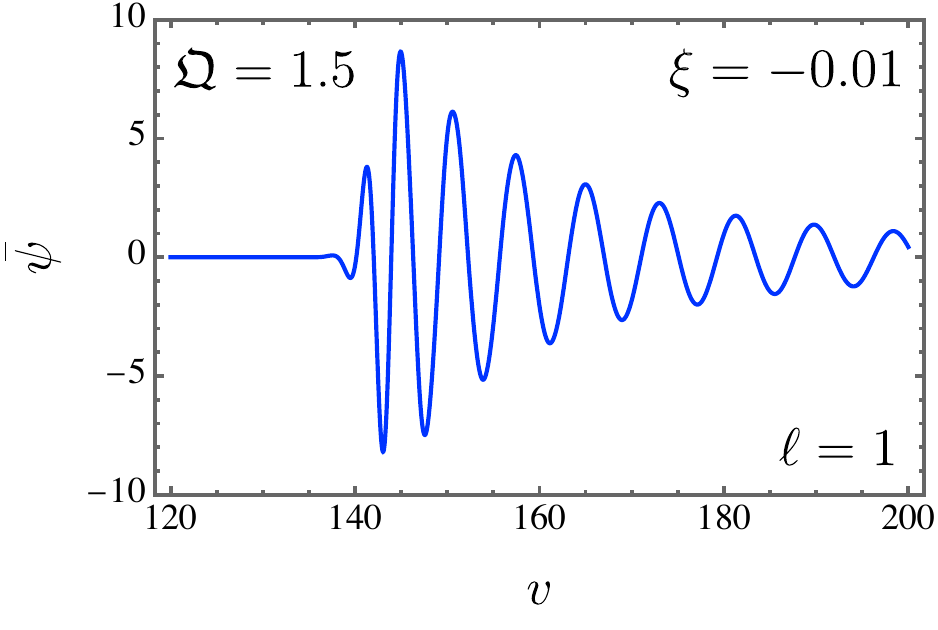}
     \caption{Numerical evolution of the waveform \(\bar{\psi}\) for scalar perturbations under varying \(\xi\) values, with parameters \(\ell = 1\), \(\mathfrak{Q} = 1.5\), and \(\Lambda = 10^{-5}\).}
    \label{psibarfunctionl1}
\end{figure}

\begin{figure}
    \centering
    \includegraphics[scale=0.45]{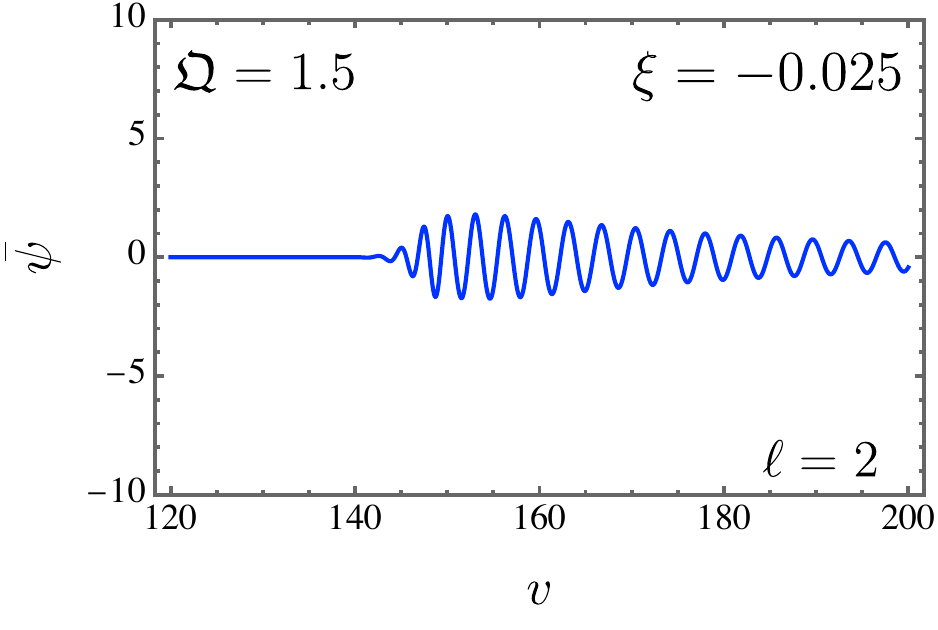}
    \includegraphics[scale=0.45]{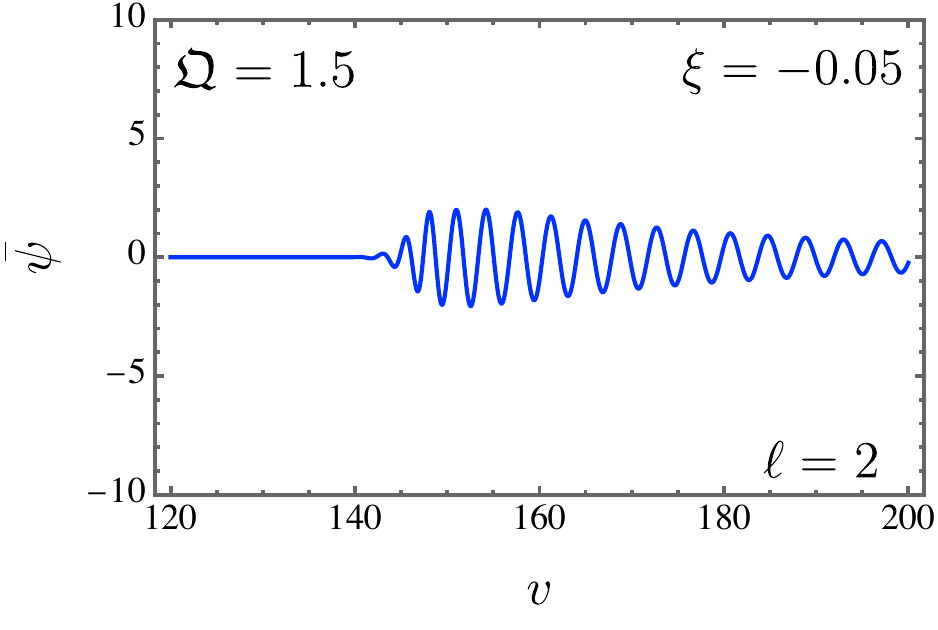}
    \includegraphics[scale=0.45]{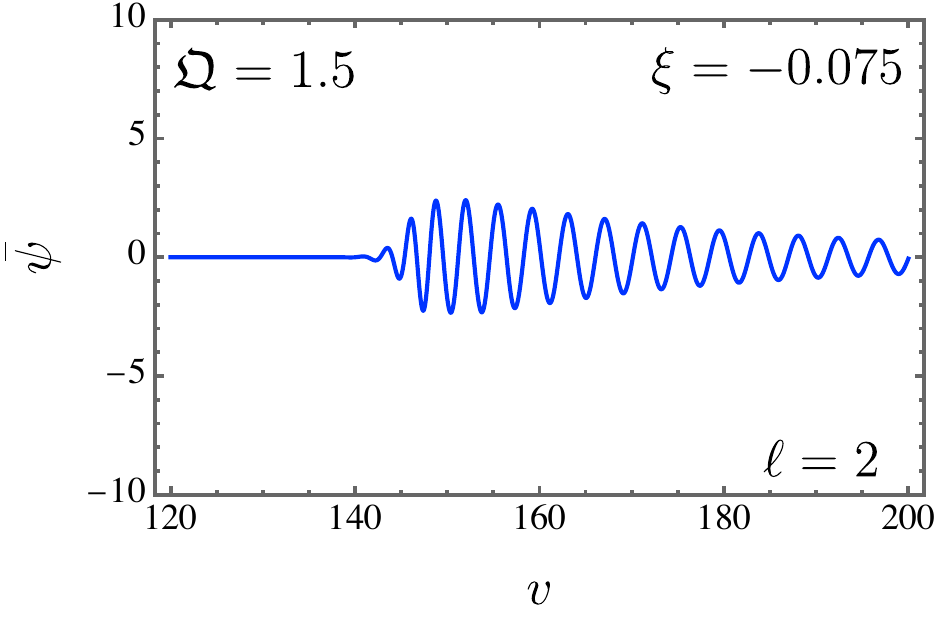}
    \includegraphics[scale=0.45]{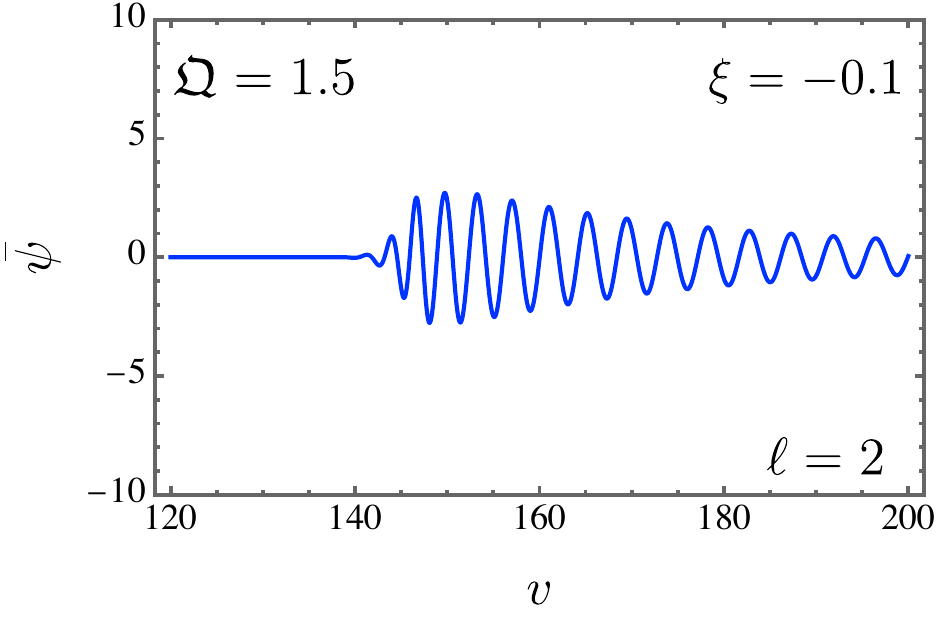}
     \caption{Numerical evolution of the waveform \(\bar{\psi}\) for scalar perturbations under varying \(\xi\) values, with parameters \(\ell = 2\), \(\mathfrak{Q} = 1.5\), and \(\Lambda = 10^{-5}\).}
    \label{psibarfunctionl2}
\end{figure}

\begin{figure}
    \centering
    \includegraphics[scale=0.45]{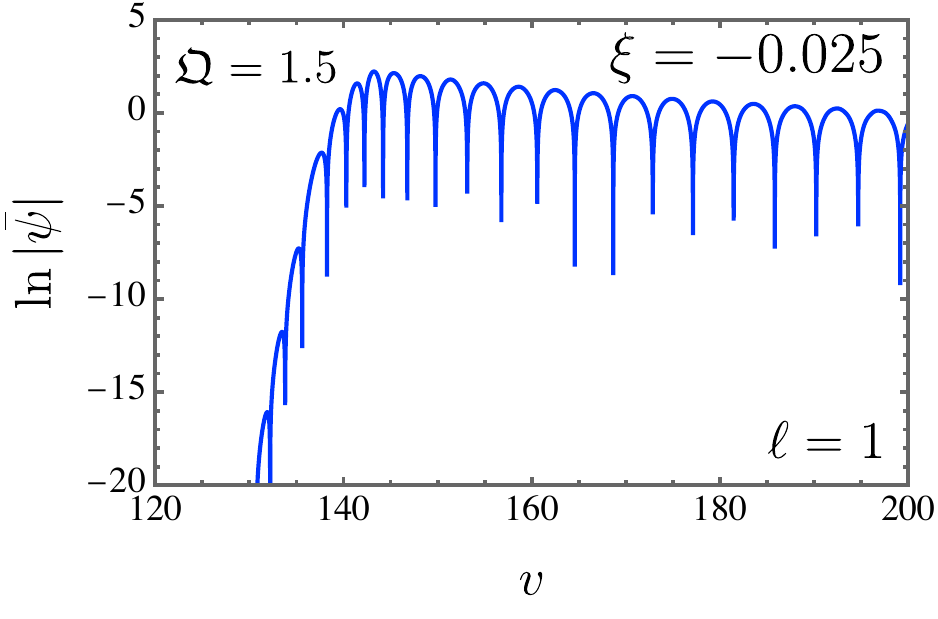}
    \includegraphics[scale=0.45]{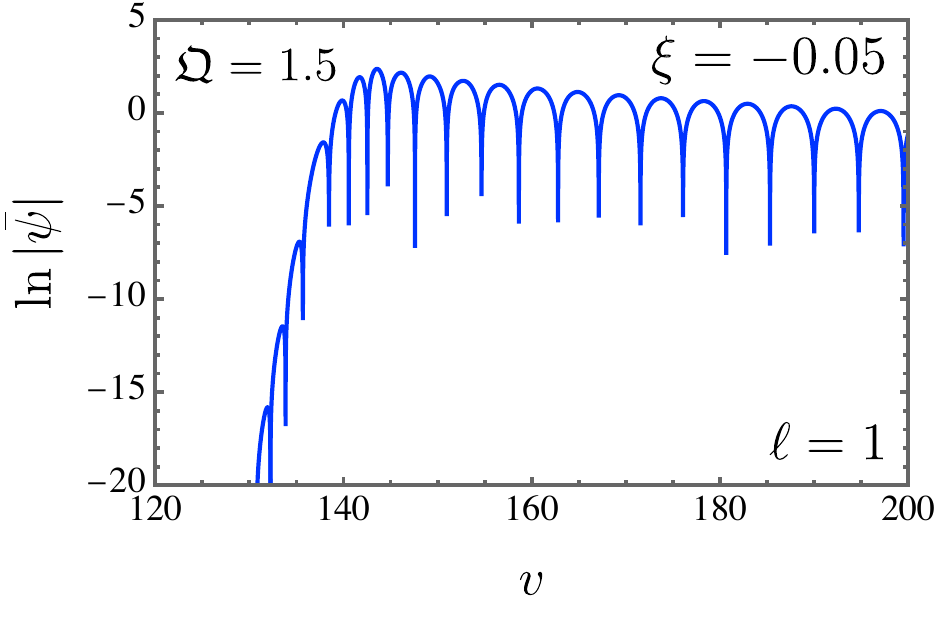}
    \includegraphics[scale=0.45]{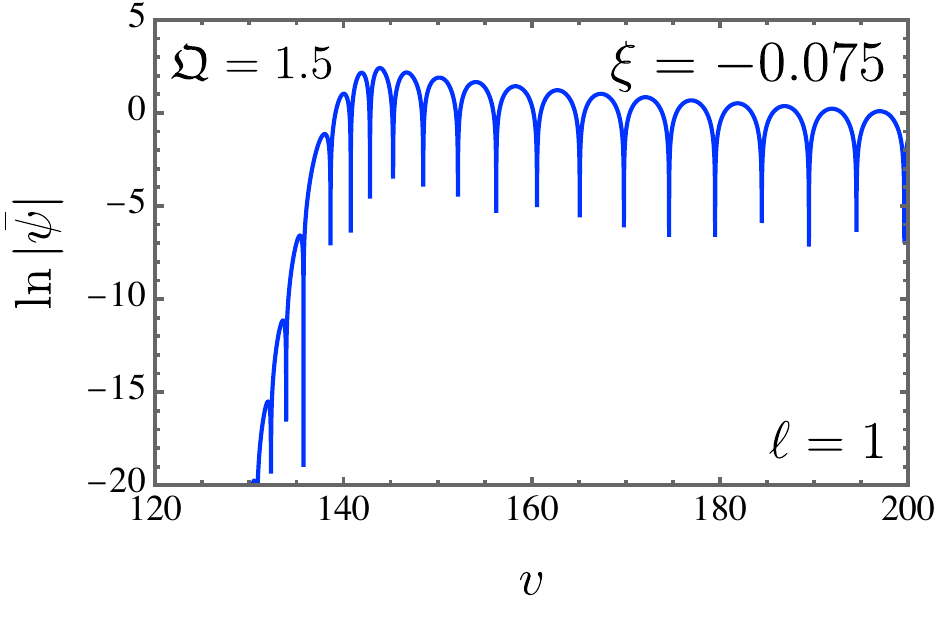}
    \includegraphics[scale=0.45]{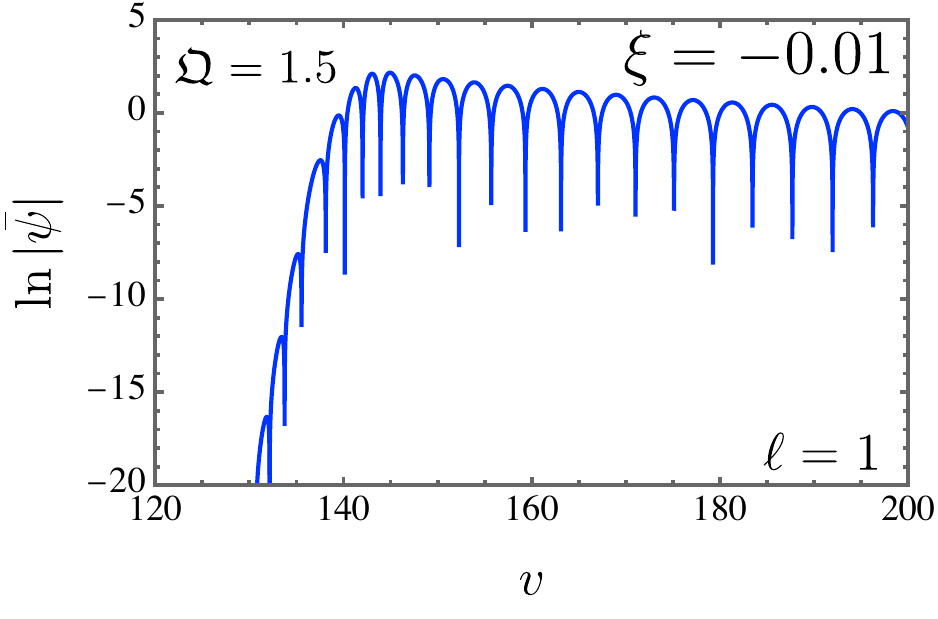}
    \caption{Numerical $\ln$ plots of the time domain profiles \(\bar{\psi}\) for scalar perturbations under varying \(\xi\) values, with parameters \(\ell = 1\), \(\mathfrak{Q} = 1.5\), and \(\Lambda = 10^{-5}\).}
    \label{logpsibarfunctionl1}
\end{figure}

\begin{figure}
    \centering
    \includegraphics[scale=0.45]{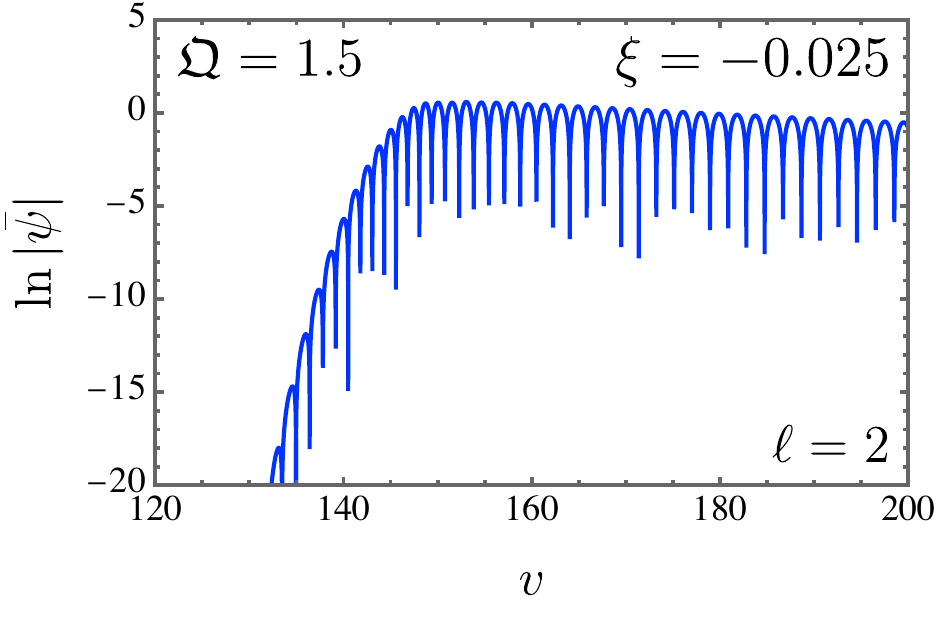}
     \includegraphics[scale=0.45]{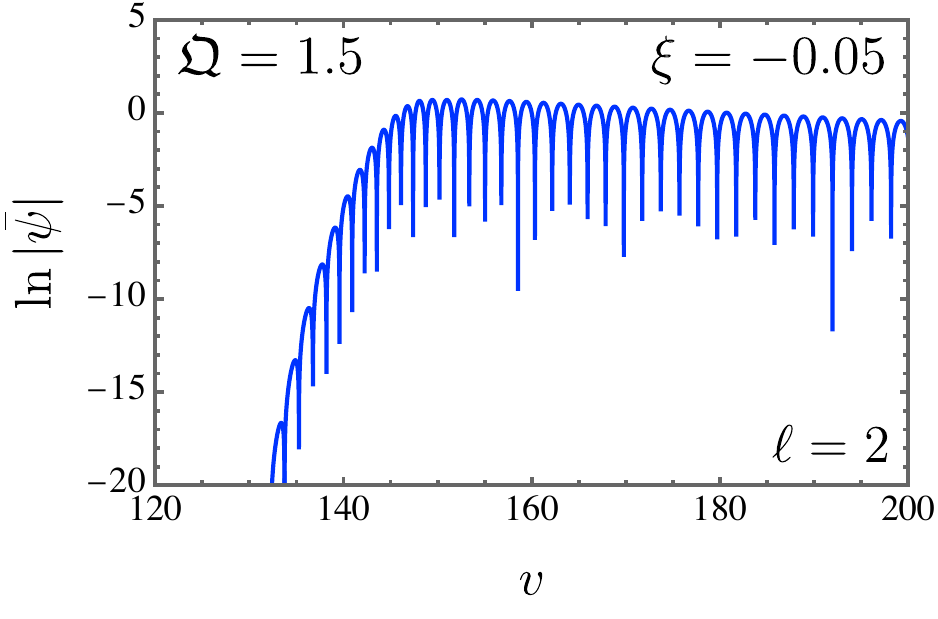}
     \includegraphics[scale=0.45]{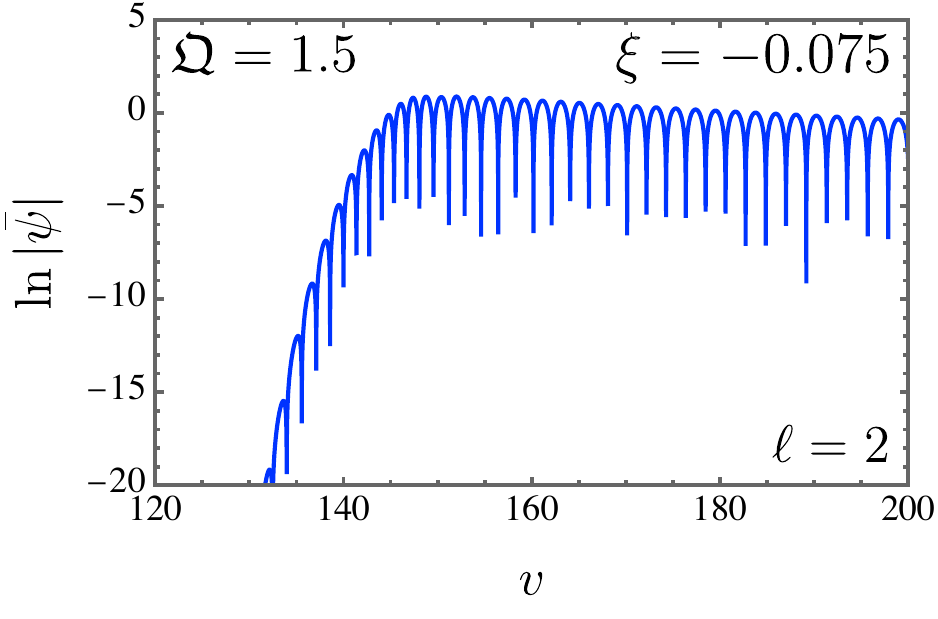}
     \includegraphics[scale=0.45]{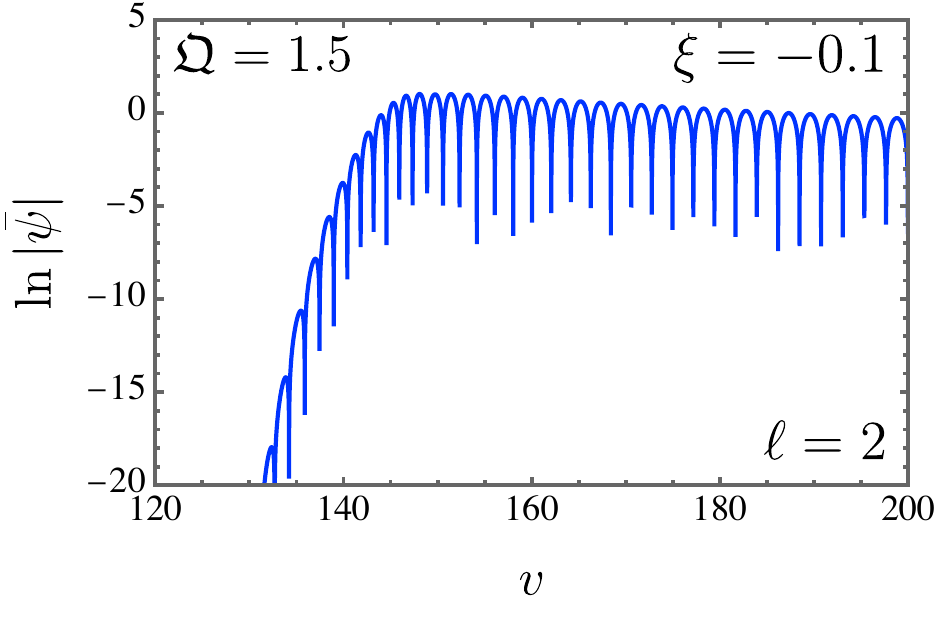}
    \caption{Numerical $\ln$ plots of the time domain profiles \(\bar{\psi}\) for scalar perturbations under varying \(\xi\) values, with parameters \(\ell = 2\), \(\mathfrak{Q} = 1.5\), and \(\Lambda = 10^{-5}\).}
    \label{logpsibarfunctionl2}
\end{figure}

Figs. \ref{psibarfunctionl1} and \ref{psibarfunctionl2} display the numerical evolution of the waveform \(\bar{\psi}\) under scalar perturbations for \(\ell = 1\) and \(\ell = 2\), respectively, with varying \(\xi\) values. The parameters are set to \(\mathfrak{Q} = 1.5\) and \(\Lambda = 10^{-5}\). Additionally, the time--domain profiles for \(\ell = 1\) and \(\ell = 2\) are shown in Figs. \ref{logpsibarfunctionl1} and \ref{logpsibarfunctionl2}, respectively, for varying \(\xi\) values. The parameters remain fixed at \(\mathfrak{Q} = 1.5\) and \(\Lambda = 10^{-5}\).

\section{Conclusion}

This work focused on investigating a nonlinear extension of the AdS Reissner--Nordström black hole. Our initial analysis examined the metric function \(f(r)\), the horizon structure, and the Ricci and Kretschmann scalars, confirming the presence of a singularity as \(r \to 0\). Additionally, we identified the non--vanishing Christoffel symbol components, which allowed us to calculate the null geodesics and analyze the impact of light paths on the photon sphere, and shadow formation. The cosmological constant \(\Lambda\) had no significant effect on the photon sphere; however, a decrease in \(\xi\) (with \(\mathfrak{Q} = 0.5\) fixed) and an increase in \(\mathfrak{Q}\) (with \(\xi = -0.1\) fixed) for \(\Lambda = 10^{-5}\) reduced the shadow radius.

Our thermodynamic analysis included an exploration of the Hawking temperature, heat capacity, and Gibbs free energy, along with an assessment of Hawking radiation. Regarding this aspect, an increase in \(\mathfrak{Q}\) and a decrease in \(\xi\) were found to reduce the particle density \(n\). These results were compared with the Schwarzschild case. We further investigated black hole evaporation by estimating its lifetime.

Quasinormal modes for scalar and vector perturbations were calculated using the WKB approximation, revealing that lower values of \(\xi\) and higher values of \(\mathfrak{Q}\) resulted in less damped oscillations. Finally, we considered the time--domain solution to evaluate the evolution of these perturbations.

Looking ahead, promising areas for exploration include detailed studies of gravitational lensing based on the extended Gauss--Bonnet theorem \cite{Li:2020wvn} and investigations into scattering phenomena and greybody factors for both fermion and boson fields. These topics, among others, are currently being actively pursued.


\section*{Acknowledgments}
\hspace{0.5cm}

A. A. Araújo Filho acknowledges the support from the Conselho Nacional de Desenvolvimento Científico e Tecnológico (CNPq) and the Fundação de Apoio à Pesquisa do Estado da Paraíba (FAPESQ) through grant [150891/2023-7]. The author extends special thanks to  N. Heidari, A. Zhidenko, A. Övgün, R. Daghigh, and M. Green for their assistance in developing the codes utilized in this manuscript. 

\section{Data Availability Statement}

Data Availability Statement: No Data associated in the manuscript


\bibliographystyle{ieeetr}
\bibliography{main}

\end{document}